\documentclass[11pt,twoside]{article}

\usepackage[
    a4paper,
    textwidth=42pc,
    textheight=53\baselineskip,
    centering,
    headheight=16pt,
    headsep=25pt,
    footskip=28pt
]{geometry}
\usepackage[mode=build]{standalone}
\usepackage[natbib=true,
    backend=biber,
    maxnames=6,
    minnames=3,
    maxcitenames=2,
    mincitenames=1,
    style=numeric,
    autocite=superscript,
    giveninits=true,
    doi=true,
    bibencoding=utf8]{biblatex}
\usepackage{url,lineno}
\usepackage{float}
\usepackage[
    font=normalsize,
    labelfont=bf,
    labelsep=period,
    justification=raggedright,
    singlelinecheck=false
]{caption}
\usepackage[onehalfspacing]{setspace}
\usepackage{array}
\usepackage{booktabs}
\usepackage{hyperref}
\hypersetup{
    colorlinks=true,
    citecolor=egyptianblue,
    linkcolor=darkcoral,   
    filecolor=black,
    urlcolor=darkolivegreen,
    bookmarksopen=true,
    bookmarksopenlevel=3,
    plainpages=false,
    pdftitle={Comparing local and global optimization reveals untapped hybrid potentials for machine learning-based inverse planning},
    pdfpagemode=FullScreen,
    }
\usepackage[all]{hypcap}
\usepackage{makecell}
\usepackage{mathtools}
\usepackage[nopatch=footnote]{microtype}
\usepackage{siunitx}
\DeclareSIUnit{\pp}{\textup{p.p.}}
\usepackage{amsmath, amsxtra, amstext, amssymb, amsthm, amsfonts}
\usepackage{dsfont}
\usepackage{tikz}
\usetikzlibrary{arrows.meta, calc, fit, patterns, positioning, shapes.geometric, shapes.symbols}
\usepackage{multirow}
\usepackage{etoolbox}
\usepackage{bigstrut}
\usepackage{authblk}
\usepackage{graphicx}
\usepackage{subcaption}
\usepackage{xcolor}
\usepackage{adjustbox}
\usepackage{cleveref}
\usepackage{csquotes}
\usepackage{fix-cm}
\usepackage{threeparttable}
\usepackage{pgfplots}
\usepackage{fancyhdr}
\usepgfplotslibrary{groupplots}
\usetikzlibrary{matrix}
\pgfplotsset{compat=1.18}
\usetikzlibrary{spy}
\usepgfplotslibrary{statistics}
\usepackage{forest}
\usepackage{orcidlink}

\definecolor{egyptianblue}{rgb}{0.06, 0.2, 0.65}
\definecolor{darkolivegreen}{rgb}{0.33, 0.42, 0.18}
\definecolor{darkcoral}{rgb}{0.8, 0.36, 0.27}
\definecolor{leafgreen}{RGB}{220, 242, 220}
\definecolor{darkleafgreen}{RGB}{200, 226, 200}

\forestset{
  tree_style/.style={
    for tree={
      draw, rounded corners, inner sep=6pt, font=\sffamily\footnotesize,
      edge={->, >=stealth}, align=center, l sep=1cm, s sep=0.8cm, fill=leafgreen
    },
    where n children=0{fill=darkleafgreen, font=\sffamily\footnotesize}{}
  }
}

\newcommand{\solid}{\raisebox{2pt}{\tikz{\draw[black,solid,line width = 0.9pt](0,0) -- (4mm,0);}}}
\newcommand{\dashed}{\raisebox{2pt}{\tikz{\draw[black,dashed,line width = 0.9pt](0,0) -- (4mm,0);}}}
\newcommand{\dotted}{\raisebox{2pt}{\tikz{\draw[black,dotted,line width = 0.9pt](0,0) -- (4mm,0);}}}

\newcolumntype{C}[1]{>{\centering\let\newline\\\arraybackslash\hspace{0pt}}m{#1}}
\newcolumntype{M}[1]{>{\centering\arraybackslash}m{#1}}

\renewcommand{\arraystretch}{1}


\makeatletter
\renewcommand{\paragraph}{%
  \@startsection{paragraph}{4}{\z@}%
  {8\p@ plus 1\p@}%
  {1ex plus 0.2ex}%
  {\normalfont\bfseries}%
}
\makeatother

\usepackage{adjustbox}
\usepackage{environ}
\usepackage{threeparttable}

\let\OriginalThreePartTable\threeparttable
\let\EndOriginalThreePartTable\endthreeparttable

\RenewEnviron{threeparttable}{%
  \begin{adjustbox}{max width=\linewidth,center}
    \OriginalThreePartTable
      \BODY
    \EndOriginalThreePartTable
  \end{adjustbox}
}

\newenvironment{scaletable}
    {\begin{adjustbox}{max width=\linewidth,center}}
    {\end{adjustbox}}

\begin{document}

\pagestyle{fancy}
\renewcommand{\sectionmark}[1]{\markboth{#1}{}}
\fancyhead{}
\renewcommand{\headrulewidth}{0.5pt}%
\renewcommand{\headrule}{\hbox to\headwidth{%
  \color{darkgray}\leaders\hrule height \headrulewidth\hfill}}
\fancyhead[RE]{\small\color{darkgray}\thepage \vspace{-10pt}}
\fancyhead[LE]{\small\color{darkgray} Ortkamp et al. (2026): Comparing local and global ML-based inverse planning \vspace{-10pt}}
\fancyhead[RO]{\small\color{darkgray}\nouppercase{\leftmark} \vspace{-10pt}}
\fancyhead[LO]{\small\color{darkgray}\thepage \vspace{-10pt}}

\fancyfoot{}
\renewcommand{\footrulewidth}{0.5pt}%
\renewcommand{\footrule}{\hbox to\headwidth{%
  \color{darkgray}\leaders\hrule height \footrulewidth\hfill}}
\fancyfoot[RO, LE]{\small\color{darkgray} Last edited \today}

\setlength{\headheight}{19.69997pt}

\title{\raggedright Comparing local and global optimization reveals untapped hybrid potentials for machine learning-based inverse planning}


\renewcommand\Authfont{\raggedright\normalsize}
\renewcommand\Affilfont{\raggedright\itshape\footnotesize}

\author[1,2,3,*]{Tim Ortkamp}
\author[2,4]{Florian Leininger}
\author[5]{Zhongxing Liao}
\author[2,4,*,$\dagger$]{Niklas Wahl}
\author[6,*,$\dagger$]{Ali Ajdari}
\affil[1]{Karlsruhe Institute of Technology (KIT), Scientific Computing Center, Hermann-von-Helmholtz-Platz 1, 76344 Eggenstein-Leopoldshafen, Germany}
\affil[2]{Department of Medical Physics in Radiation Oncology, German Cancer Research Center (DKFZ), Im Neuenheimer Feld 280, 69120 Heidelberg, Germany}
\affil[3]{Helmholtz Information and Data Science School for Health (HIDSS4Health), Karlsruhe/Heidelberg, Germany}
\affil[4]{National Center for Radiation
Research in Oncology (NCRO), Heidelberg Institute for Radiation Oncology (HIRO), Im Neuenheimer Feld 280, 69120 Heidelberg, Germany}
\affil[5]{Department of Thoracic Radiation Oncology, Division of Radiation Oncology, University of Texas' MD Anderson Cancer Center, Houston, TX, USA}
\affil[6]{Department of Radiation Oncology, Mass General Brigham Cancer Institute \& Harvard Medical School, Boston, MA, USA}

\date{\phantom{}\\[-2.8em] \begin{minipage}{\textwidth}\footnotesize{$^\dagger$These authors share senior authorship. \\[1.5em] Correspondence*: \\[0.3em] Tim Ortkamp \\[-0.3em] tim.ortkamp@kit.edu \\[0.3em] Niklas Wahl \\[-0.3em] n.wahl@dkfz.de \\[0.3em] Ali Ajdari \\[-0.3em] aajdari@mgh.harvard.edu} \\[1.3em] \normalsize Version typeset \today \end{minipage}}

\maketitle
\thispagestyle{empty}
\renewcommand{\thefootnote}{\arabic{footnote}}


\begin{abstract}
\label{sec:abstract}
\noindent
\textbf{Background:} Machine learning (ML) models for normal tissue complication probability (NTCP) can be embedded directly into inverse treatment planning, but the resulting optimization problems are generally nonconvex, and the choice between global and local optimization involves a trade-off between optimality guarantees and computational cost.

\noindent
\textbf{Aims:} Two open-source frameworks for ML-based inverse planning were compared: OptiCL (global mixed-integer optimization via constraint learning) and pyanno4rt (local nonlinear optimization via end-to-end differentiation). Furthermore, a proof-of-concept was provided for hybrid strategies that transfer information from the local to the global solver.

\noindent
\textbf{Methods:} Logistic regression, linear support vector machine, and feed-forward neural network NTCP models for grade 2+ xerostomia and grade 2+ radiation pneumonitis were embedded. Plan optimization was performed for eight head-and-neck (HN) photon cases and seven lung proton cases, yielding 120 treatment plans across both frameworks. These plans were compared by target coverage, OAR dose metrics, spatial dose agreement, post-hoc predicted NTCP, and wall-clock runtime, using patient-level paired differences. A projection-based gradient approximation was introduced to embed decision trees into gradient-based optimization. In addition to that, warm-starting, constraint strengthening, and pruning were investigated as hybrid strategies for combining local and global optimization.

\noindent
\textbf{Results:} Comparing OptiCL and pyanno4rt across the HN and lung treatment plans revealed broadly consistent dose distributions. Mean differences ($\text{OptiCL}-\text{pyanno4rt}$ [\SI{95}{\percent} CI]) were small: for HN cases, \SI{-0.16}{\gray} $[-0.37, 0.06]$ (parotid) and \SI{-0.00}{\gray} $[-0.01, 0.00]$ (PTV); for lung cases, \SI{-0.52}{\gray} $[-1.10, 0.06]$ (total lung) and \SI{-0.04}{\gray} $[-0.07, -0.01]$ (PTV). Mean NTCP differences amounted to \SI{-0.08}{\pp} $[-0.49, 0.33]$ (HN) and \SI{-1.73}{\pp} $[-2.68, -0.77]$ (lung). OptiCL tended toward lower OAR doses and predicted NTCP, but required one to two orders of magnitude higher wall-clock runtime. Hybridization of the decision tree-based HN plan eliminated \SI{80}{\percent} of binary decision variables without compromising plan quality.

\noindent
\textbf{Conclusion:} Global constraint learning and local nonlinear optimization yield broadly consistent NTCP-based treatment plans, with complementary strengths in solution quality and computational efficiency. While OptiCL prioritizes global optimality, pyanno4rt offers substantially shorter runtimes. Hybridization can reduce the complexity of global mixed-integer formulations, highlighting the potential for combining both paradigms in ML-based inverse planning. \\[0em]

\scriptsize
\noindent
\textbf{Keywords:} outcome-driven inverse planning, normal tissue complication probability (NTCP), machine learning, mixed-integer programming (MIP), nonlinear programming (NLP), hybrid optimization strategies, xerostomia, radiopneumonitis
\end{abstract}

\section{Introduction} \label{sec:introduction}
Inverse planning has traditionally relied on physical dose constraints and dose-volume histograms (DVHs) as surrogate measures for biological outcomes. This detaches the optimization process from the underlying clinical balance between tumor control probability (TCP) and normal tissue complication probability (NTCP), formulating it as a dosimetric trade-off: \enquote{How can I maximize the physical dose to the tumor while minimizing it for organs-at-risk (OARs)?}. Consequently, optimization mostly relies on dose surrogate objectives---such as quadratic deviations from empirical dose prescriptions---in which dose-volume objectives and population-based tolerance doses stand in for individual complication risk, rather than directly minimizing the predicted toxicity for the individual patient. As a result, plans that satisfy all dose-volume criteria may still be suboptimal with respect to individual complication risk, while personal risk factors and their interplay with dose remain unaccounted for during optimization.

Recent advancements in machine learning (ML) for TCP and NTCP modeling, e.g. \cite{Abdollahi2019,Luna2019,Men2019,Jalalifar2022,Chamseddine2023} for specific modeling approaches and \cite{Isaksson2020,Appelt2022} for systematic reviews, have catalyzed a definitive shift toward outcome-driven inverse planning \cite{Nahum2012,Kierkels2014,Ajdari2022} which asks the more pertinent question: \enquote{How can I maximize tumor control while minimizing the occurrence of post-treatment complications?}. ML architectures have already selectively demonstrated superiority over classical radiotherapy treatment outcome models \cite{Dijk2017,Dijk2018,Rossi2018,Gabrys2018,Luna2019}, not only for reasons of outperforming them in terms of predictive quality, but also thanks to their capacity to synthesize high-dimensional and heterogeneous inputs \cite{Kang2015}. Such inputs range from dosiomics \cite{Placidi2021}, radiomics \cite{Dijk2017,Dijk2018}, demographics and clinical variables to potentially complex and large-scale multi-omics data \cite{Kang2020}. By inferring the complex, nonlinear relationships between patient features and treatment outcomes, these models facilitate \enquote{high-fidelity} personalized planning tailored to a specific patient's risk profile.

Moving from physical surrogates to NTCP-driven objectives, however, also requires a new perspective on optimization. Conventional optimizers and treatment planning systems (TPS) can accommodate individualized planning as long as the NTCP model is additive in its dosimetric and non-dosimetric inputs and relies on a single dominant dose metric; in this case, the contribution of the patient-specific covariates can be factorized out and translated into a personalized dose constraint. This approach reaches its limits, however, when the model is not linearly separable or when several dosimetric predictors enter simultaneously, so that the trade-off between them is defined by the model rather than by a single tolerance value. 

To fully leverage ML-based NTCP models in inverse planning, the prediction function should therefore be embedded directly within the optimization model. In this setting, the optimizer evaluates the NTCP prediction for every candidate dose distribution, with the patient-specific covariates (e.g., demographics or radiomic shape descriptors) held fixed at the values of the individual patient. Importantly, this does not require any assumption about how dosimetric and non-dosimetric inputs interact within the model, allowing the full prediction model to be incorporated without having to derive patient-specific dose constraints for each model structure. Depending on the model and its dose-dependent features, the resulting optimization problem may be nonconvex in the fluence, motivating different strategies for solving the embedded prediction problem. 

\citet{Maragno2024} propose a framework inspired by \emph{optimization by constraint learning} (OCL), which reformulates ML models as linear or mixed-integer constraints. This enables rigorous global optimality guarantees via solvers such as branch-and-bound, but at the cost of poor scalability and restricted model complexity. Conversely, the \emph{end-to-end differentiation} framework introduced by \citet{Ortkamp2024} prioritizes computational efficiency, leveraging gradient-based methods to handle even highly expressive, nonlinear models. While this approach performs well on large-scale problems, it does not generally provide global optimality guarantees, as local nonlinear optimization may terminate at stationary points that are not globally optimal. In the context of radiotherapy, this distinction has direct implications for plan quality: a local optimizer may converge to a suboptimal stationary point of a nonconvex NTCP objective, yielding a plan whose predicted complication risk exceeds that of the globally optimal solution to the same problem, without providing the planner with any indication that a better solution exists. Conversely, the computational cost of global optimization determines whether NTCP-based planning is compatible with time-sensitive clinical workflows, including interactive planning, replanning, and multi-criteria exploration. Whether, and to what extent, local and global solutions differ for realistic NTCP models is therefore an empirical question that this study seeks to address.

To this end, we investigate the trade-off between global optimality and computational efficiency in ML-based inverse planning by comparing two optimization frameworks: \emph{OptiCL} \cite{OptiCL2021}, which implements OCL, and \emph{pyanno4rt} \cite{pyanno4rt2026}, which utilizes end-to-end differentiation. Analyzing treatment plans generated across eight head-and-neck (HN) and seven lung cancer cases (120 plans total), we quantify dosimetric and NTCP-based differences between these platforms. Here, predictions for normal-tissue toxicities are incorporated into model-based optimization functions, utilizing NTCP models for grade 2+ xerostomia and radiopneumonitis. Although technically feasible, TCP models and outcome-based optimization in a broader sense are outside the scope of this study, leaving tumor control evaluated purely dosimetrically. To explore the feasibility of hybridization, i.e., the transfer of information from the local nonlinear solution (optimized fluence and predicted NTCP) to the global mixed-integer solver, we leverage the nonlinear programming (NLP) solution of a decision tree-based HN plan to inform \emph{warm-start}, \emph{strengthening}, and \emph{pruning} techniques. This approach allows to presolve the mixed-integer programming (MIP) formulation within OptiCL and reduce the computational complexity.

In the present study, we aimed to (i) compare OptiCL and pyanno4rt for NTCP-embedded treatment plan optimization on identical planning problems, considering logistic regression, linear support vector machine (SVM), and feed-forward neural network models for grade 2+ xerostomia and grade 2+ radiopneumonitis; (ii) quantify differences in dosimetry, predicted NTCP, and runtime between global mixed-integer and local nonlinear optimization across eight head-and-neck photon cases and seven lung proton cases; and (iii) investigate whether hybridization strategies that transfer information from the local to the global solver can reduce the complexity of the mixed-integer formulation without compromising plan quality. We hypothesized that both frameworks would yield consistent plans for convex (conventional) optimization problems, that differences would arise with nonconvex ML components, and that the local solution could provide useful a priori information to guide the global solver. The methodological contributions comprise a projection-based gradient approximation that enables decision trees to be optimized using gradient-based methods, a reusable interface for translating pyanno4rt problems into OptiCL formulations, and a proof-of-concept hybridization strategy applied to a decision tree-based planning problem.

\section{Materials and methods} \label{sec:methods}

\subsection{Inverse planning with machine learning NTCP models} \label{subsec:inverse-planning}
This section formalizes the integration of the ML-based NTCP models within the inverse planning framework. We first define the radiotherapy optimization model and then formulate the prediction functions for logistic regression, linear SVMs, feed-forward neural networks, and decision trees.

\subsubsection{Optimization model} \label{subsubsec:optimization-model}
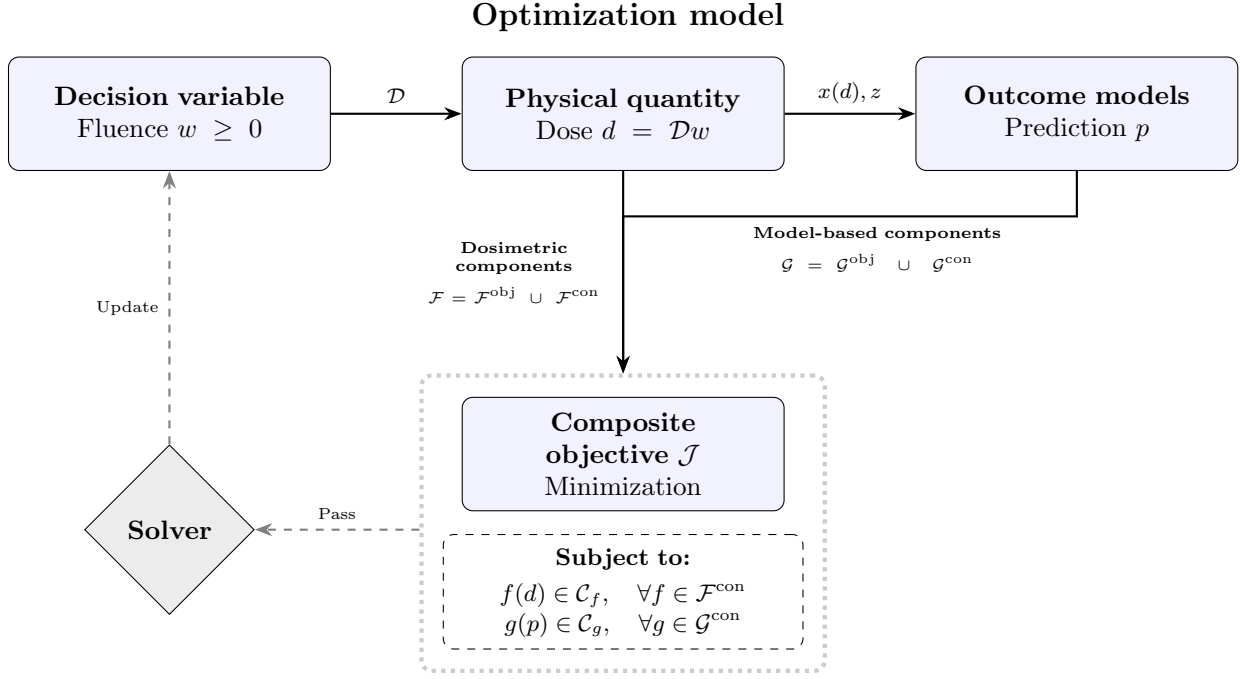
\begin{figure}[!htbp]
    \centering
    {\large\textbf{Optimization model}} \\[1.8ex]
    \begin{tikzpicture}
    [
    block/.style={rectangle, draw, fill=blue!5, text width=4cm, align=center, minimum height=1.5cm, rounded corners, font=\small},
    math/.style={rectangle, draw, dashed, fill=white, text width=4.5cm, align=center, minimum height=1.5cm, rounded corners, font=\footnotesize},
    solver/.style={diamond, draw, fill=gray!15, text width=2cm, align=center, font=\small\bfseries, inner sep=0pt},
    arrow/.style={-Stealth, thick},
    ]

    \node [block] (phi) at (0,0) {\textbf{Decision variable} \\ Fluence $w \geq 0$};
    \node [block] (dose) at (6,0) {\textbf{Physical quantity} \\ Dose $d = \mathcal{D}w$};
    \node [block] (preds) at (12,0) {\textbf{Outcome models} \\ Prediction $p$};
    
    \node [block] (obj) at (6,-4.5) {\textbf{Composite objective $\mathcal{J}$} \\ Minimization};

    \node [math, anchor=north] (cons_text) at (obj.south) [yshift=-0.3cm] {
        \begin{tabular}{c}
            \textbf{Subject to:} \\[0.3em]
            $f(d) \in \mathcal{C}_{f}, \quad \forall f \in \mathcal{F}^{\text{con}}$ \\
            $g(p) \in \mathcal{C}_{g}, \quad \forall g \in \mathcal{G}^{\text{con}}$
        \end{tabular}
    };

    \node [solver] (solver) at (0,-5.5) {Solver};

    \node [draw, black!20, dotted, ultra thick, rounded corners, inner sep=8pt, fit=(obj) (cons_text)] (functions) {};

    \draw [arrow] (phi) -- (dose) node[midway, above, font=\scriptsize] {$\mathcal{D}$};
    \draw [arrow] (dose) -- (preds) node[midway, above, font=\scriptsize] {$x(d), z$};
    
    \draw [arrow] (preds.south) -- ++(0,-0.6) -| (functions.north) 
        node[pos=0.22, below, font=\tiny, text width=4cm, align=center] {
            \textbf{Model-based components} \\[0.5em] $\mathcal{G}=\mathcal{G}^{\text{obj}}\;\cup\;\mathcal{G}^{\text{con}}$
        };
    
    \draw [arrow] (dose.south) -- (functions.north) 
        node[midway, left, font=\tiny, text width=2.6cm, align=center] {
            \textbf{Dosimetric components} \\[0.5em] $\mathcal{F}=\mathcal{F}^{\text{obj}}\;\cup\;\mathcal{F}^{\text{con}}$
        };

    \draw [arrow, dashed, gray] (functions.west |- solver) -- (solver.east)
        node[midway, above, font=\tiny, black] {Pass};

    \draw [arrow, dashed, gray] (solver.north) -- (phi.south)
        node[midway, left, font=\tiny, black, align=right] {Update};

\end{tikzpicture}
    \caption{\textbf{Schematic of the optimization model}. The workflow maps the fluence $w$ (decision variable) to the physical dose $d$ from which input features $x(d)$ are extracted for NTCP prediction. Both dosimetric components $f$ and model-based components $g$ are then integrated into a composite objective function $\mathcal{J}$ and constraint sets $\mathcal{C}_{f}$ and $\mathcal{C}_{g}$ for iterative optimization via the chosen solver.}
    \label{fig:optimization-model}
\end{figure}

The treatment planning task is represented as a unified constrained optimization problem, integrating conventional dosimetric criteria and NTCP model-based predictions. In terms of physics, we define the dose-influence matrix $\mathcal{D}$ and the beamlet intensities $w$ (the primary decision variables), which map linearly to the physical dose distribution $d=\mathcal{D}w$, where $w\geq{0}$. Regarding the NTCP models, let $\mathbb{P}$ denote the predictive mapping that relates a dose-derived feature vector $x$ and constant patient-specific covariates $z$ to an individual NTCP prediction $p$. Structurally, we distinguish two types of clinical criteria: dosimetric functions $f\in{\mathcal{F}}$ (e.g., dose-volume or EUD-based metrics) and model-based functions $g\in{\mathcal{G}}$ evaluated on the NTCP measure $p$. The latter allow for transformations of $p$, such as negation for TCP-related mappings $\mathcal{P}$ (notwithstanding their omission in this study). Depending on the clinical strategy, each criterion is either assigned as a planning objective $f\in{\mathcal{F}^{\text{obj}}}$ or $g\in{\mathcal{G}^{\text{obj}}}$ to be minimized, or as a constraint $f\in{\mathcal{F}^{\text{con}}}$ or $g\in{\mathcal{G}^{\text{con}}}$ to be strictly respected. Furthermore, we use $\mathcal{J}$ as the composite objective function---such as a weighted-sum formulation---to aggregate all individual objectives in $\mathcal{F}^{\text{obj}}$ and $\mathcal{G}^{\text{obj}}$ according to the planning priorities. The feasible regions for the dosimetric and model-based constraints are defined by $\mathcal{C}_{f}$ and $\mathcal{C}_{g}$, respectively. Therefore, we formally express the optimization model in its most general form, as depicted in \Cref{fig:optimization-model}:
\begin{equation} \label{eq:optimization-model}
    \begin{aligned}
        & \underset{w \geq{0}}{\text{minimize}} & & \mathcal{J} \left( \mathcal{F}^{\text{obj}},\mathcal{G}^{\text{obj}} \right) \\
        & \text{subject to} & & f(d) \in \mathcal{C}_{f}, \quad \forall f\in\mathcal{F}^{\text{con}} \\
        & & & g(p) \in \mathcal{C}_{g}, \quad \forall g\in\mathcal{G}^{\text{con}} \\
        & & & p = \mathbb{P}(x,z) \\
        & & & x = x(d) \\
        & & & d = \mathcal{D}w\,.
    \end{aligned}
\end{equation}
For inverse planning, formulation (\ref{eq:optimization-model}) constitutes a large-scale constrained problem, generally requiring MIP or NLP techniques. The computational tractability and the assurance of global optimality are governed by the model's mathematical structure---specifically its convexity, a property frequently violated by the inherently nonconvex nature of many machine learning architectures.

\subsubsection{NTCP model embeddings} \label{subsubsec:model-embedding}
For embedding NTCP predictions into the optimization model from formulation (\ref{eq:optimization-model}), we utilize a suite of ML algorithms, including logistic regression, linear SVMs, feed-forward neural networks, and decision trees. These algorithms induce the predictive mapping function $\mathbb{P}$ with parameters learned from historical toxicity data. This section formally introduces the mathematical representations of these mappings, drawing on the formulations established in our previous work \citep{Maragno2024,Ortkamp2024}.

\paragraph*{Logistic regression, linear support vector machines, and feed-forward neural networks}
For the first three algorithms, we closely follow standard formulations. Let $(x,z)$ denote the input features, $\eta$ the linear predictor, and $\sigma$ the sigmoid function. Logistic regression maps $(x,z)$ to $p$ via
\begin{align} \label{eq:log-reg-pred}
    \mathbb{P}_{LR}(x,z)=\sigma(\eta) \quad \text{where } \eta=\beta_{x}^{T}x+\beta_{z}^{T}z+c\,.
\end{align}
Similarly, linear SVMs yield a sigmoidal mapping via the Platt scaling function $\sigma_{P}$ \cite{Platt2000}, i.e.,
\begin{align} \label{eq:svm-pred}
    \mathbb{P}_{SVM}(x,z)=\sigma_{P}(\eta) \quad \text{where } \eta=\omega_{x}^{T}x+\omega_{z}^{T}z+c\,.
\end{align}
Feed-forward neural networks with layer-wise weight matrices $W_{l}$, biases $b_{l}$, and activation functions $a_{l}$, for $l=1,2,\ldots,L-1$, produce their final output as follows:
\begin{equation} \label{eq:nn-pred}
    \begin{aligned}
        \mathbb{P}_{NN}(x,z) &= \sigma\big(W_{L}h_{L-1}+b_{L}\cdot{\mathbf{1}}\big)\,, \\ h_{l} &= a_{l}\big(W_{l}h_{l-1}+b_{l}\cdot{\mathbf{1}}\big)\,, \quad l=1, \ldots, L-1\,, \\ h_{0}&=(x,z)\,.
    \end{aligned}
\end{equation}
\citet{Maragno2024} provide the constraint-based formulations underlying this work, which incorporate a threshold $\tau$ on the predicted NTCP. Specifically, they use ReLU activation functions in neural networks to enable a big-$M$ approach that exactly represents the network's logic through linear constraints.

\paragraph*{Decision trees}
To integrate decision trees into optimization, \citet{Maragno2024} further propose the following strategies: either decomposing the problem into parallel subproblems restricted to individual leaves that satisfy a probability threshold $\tau$, or formulating a monolithic mixed-integer program that explicitly encodes the tree structure via binary variables. Despite the mathematical rigor of these discrete formulations, the resulting combinatorial complexity often impairs performance, particularly as the dimensionality of the feature space increases. To circumvent these scaling issues, we introduce a gradient-based approximation designed to navigate the tree's step-function landscape as follows:

\begin{enumerate}
    \item \textit{Path set construction:} Represent the tree as a set of all root-to-leaf paths $\mathcal{S}$ with pairs of leaf probabilities and corresponding polyhedral regions.
    \item \textit{Directional filtering:} At each iteration $t$, evaluate the current prediction $p^{(t)}=\mathbb{P}(x^{(t)}, z)$ for the dosimetric feature vector $x^{(t)}$. Then, directions for improvement can be derived from the set of \enquote{non-inferior} paths $\tilde{\mathcal{S}}\subseteq{\mathcal{S}}$ where the leaf probability is smaller or equal to $p^{(t)}$.
    \item \textit{Feasible union projection:} Determine the next iterate $x^{(t+1)}$ by projecting onto the union of the polyhedral regions (constraint sets) associated with $\tilde{\mathcal{S}}$.
    \item \textit{Surrogate gradient estimation:} Approximate the gradient as the normalized difference between the current point and the nearest point on a superior plateau, scaled by the change in the objective value:
    \begin{equation} \label{eq:tree-grad}
        \nabla_{x^{(t)}} \mathbb{P} \approx \frac{(x^{(t+1)} - x^{(t)})}{||x^{(t+1)} - x^{(t)}||_{2}^2} \left( \mathbb{P}(x^{(t+1)}, z) - p^{(t)} \right).
    \end{equation}
    If $x^{(t+1)}=x^{(t)}$ or $\tilde{\mathcal{S}}$ is empty, the gradient is set to zero.
    \item \textit{Iteration:} Repeat steps 2--4 until convergence.
\end{enumerate}

\subsection{Hybridization strategies} \label{subsec:hybrid-strategies}
Building upon established MIP and NLP foundations, we apply different strategies for combining these optimization paradigms to address the inherent trade-off between accuracy and scalability in ML-based inverse planning. In particular, we detail three hybridization strategies that leverage off-the-shelf solvers, obviating the need for a novel hybrid algorithm: \emph{primal warm-starts}, \emph{strengthening}, and \emph{pruning}.

\subsubsection{Primal warm-starts} \label{subsubsec:warm-start}
Locally optimal solutions $w^{\ast}_{\text{NLP}}$ obtained from an NLP solver are utilized to warm-start the MIP solver by setting the initial incumbent $w^{(0)}_{\text{MIP}}=w^{\ast}_{\text{NLP}}$. This provides the MIP solver with a high-quality incumbent solution, which immediately \emph{tightens the upper bound} (for minimization problems). Potential benefits are: (i) more efficient pruning of the search tree, as nodes with a lower bound (dual bound) worse than the warm-start solution can be discarded early; (ii) improved heuristic performance, as solvers can leverage the local search in the neighborhood of the warm-start solution; and (iii) enhanced numerical stability, as the warm-start solution ensures a physically consistent starting point for the feasibility-seeking phase.

\subsubsection{Strengthening} \label{subsubsec:strengthening}
From a locally optimal solution $w^{\ast}_{\text{NLP}}$, the individual NTCP prediction $p^{\ast}$ is obtained via
\begin{align}
    p^{\ast} = \mathbb{P}\left(x(d(w^{\ast}_{\text{NLP}})), z\right)\,.
\end{align}
This value provides a valid upper bound on the globally achievable NTCP for the respective model, which we exploit to \enquote{strengthen} a model-based constraint $g$ by restricting its feasible region $\mathcal{C}_{g}$ to a subset $\tilde{\mathcal{C}}_{g}\subseteq{\mathcal{C}_{g}}$. For instance, if the original formulation includes an active NTCP constraint $g\in{\mathcal{C}_{g}=[0, 0.1]}$, and the NLP solver identifies a solution with $p^{\ast}=0.05$, the constraint $\mathcal{C}_{g}$ is tightened to $\tilde{\mathcal{C}}_{g}=[0, 0.05]$. This strategy effectively \emph{elevates the lower bound} (for minimization problems) by truncating the search space of the continuous relaxation, thereby enabling earlier pruning of suboptimal branches during the exploration of the search tree.

\subsubsection{Pruning} \label{subsubsec:pruning}
To reduce the combinatorial complexity of MIP formulations, pruning physically removes components (e.g. nodes, constraints, ...) \emph{prior} to constructing the optimization model. This structural reduction prevents the solver from exploring irrelevant branches of the search space. For decision trees, we efficiently implemented this strategy by migrating the path filtering step from \Cref{subsubsec:model-embedding}, thereby ensuring that constraints are only generated for leaves within the subset $\tilde{\mathcal{S}}$.

\subsection{Implementation} \label{subsec:implementation}
The proposed methodology is implemented using open-source optimization frameworks. This section describes the software environment and the interface developed to facilitate their comparison/hybridization.

\subsubsection{Software} \label{subsubsec:software}
Our implementation utilizes two primary Python-based frameworks: (i) OptiCL v0.0.1 \cite{OptiCL2021,Maragno2023} and (ii) pyanno4rt \enquote{Amadeus} v1.0.0 \cite{Ortkamp2024,pyanno4rt2026}. For dose calculation, we resorted to using the fast pencil-beam algorithm with voxel raytracing \cite{Siddon1985,Hong1996} as implemented in the matRad toolkit \cite{Wieser2017}. 

Numerical optimization is driven by the Gurobi Optimizer v13.0.1 \cite{Gurobi2026} within OptiCL, using a primal-dual interior-point method with a logarithmic barrier function for continuous relaxations and a parallel branch-and-bound framework for MIP. This solver is configured with a relative MIP optimality gap of $10^{-3}$, feasibility and optimality tolerances of $10^{-7}$, and the barrier method for solving the continuous subproblems at the branch-and-bound nodes. Within pyanno4rt, continuous optimization problems are solved using the interior-point algorithm provided by ipyopt v0.12.10 \cite{Ipyopt2021}. In this case, we configured ipyopt with a convergence tolerance of $10^{-8}$, constraint violation and dual infeasibility tolerances of $10^{-7}$, an adaptive barrier parameter strategy, and a limited-memory BFGS Hessian approximation with a history size of 50. NLP scaling is disabled, and the objective function change tolerance is set to $10^{-4}$. 

Computations were conducted on three hardware configurations: (i) a Windows 11 workstation equipped with an Intel Core i7-10700K CPU (3.80 GHz, 16 logical processors), 128 GB RAM, and an NVIDIA RTX A5000 GPU (24 GB VRAM), (ii) a Windows 11 mobile workstation featuring an Intel Core Ultra 9 285HX (2.80 GHz, 24 logical processors) with 32 GB RAM, and (iii) a mobile workstation running Ubuntu 20.04.6 LTS with an Intel Core i7-8565U CPU (1.80 GHz, 8 logical processors) and 16 GB RAM. To distribute the computational workload while maintaining consistency across the optimization and modeling experiments, the machines were assigned distinct tasks: (i) handled computationally intensive optimizations across all treatment plans, (ii) was used for post-optimization treatment plan analysis and plotting, and (iii) was used to fit the ML-based NTCP models. To limit computational costs, a time limit of 24 hours was imposed on each optimization run, particularly constraining the more complex MIP formulations.

\subsubsection{Hybridization} \label{subsubsec:hybridization}
To facilitate the comparison of treatment plans and the application of hybrid strategies, we implemented a Python interface class that translates pyanno4rt optimization models into Gurobi-compatible formulations within OptiCL. This process involves mapping objective functions, constraints, and input feature definitions into Gurobi's mathematical representation. Specifically, the interface handles the conversion of Keras models into Scikit-learn MLP formats, configures the constraint learner, (optionally) \enquote{injects} the NLP outputs, and executes the solver. Additionally, the interface generates comprehensive snapshots of the optimization state to ensure reproducibility and enable post-optimization analysis.

\subsection{Data} \label{subsec:data}
We use two clinical cohorts, HN and lung, each serving two purposes. First, NTCP datasets containing clinical toxicity outcomes are used to train and evaluate the ML models (\Cref{subsec:ntcp-modeling}). Second, subsets of eight HN and seven lung patients with available imaging data and segmentations are selected as planning cases for comparing OptiCL and pyanno4rt (\Cref{subsec:planning-study}). \Cref{tab:datasets} summarizes the NTCP datasets.

\paragraph*{Head-and-neck}
The HN NTCP dataset was introduced by \citet{Gabrys2018} and comprises 153 anonymized patients who received IMRT at Heidelberg University Hospital (UKHD) between 2010 and 2015.\footnote{Approved by the Ethics Committee of Heidelberg University, Nr.~S-392/2016 \enquote{Validation and development of probabilistic prediction models for radiation-induced xerostomia}.} Primary tumor locations are restricted to the oropharynx and nasopharynx. Dataset curation excluded patients with nonzero baseline xerostomia, multimodal treatments, or tumors involving the parotid glands. The dataset includes xerostomia grades according to the Common Terminology Criteria for Adverse Events (CTCAE) v4.03 at three time points: early, late, and long-term. In this study, long-term xerostomia of grade 2 or higher (2+) is defined as the primary endpoint.

Among the eight HN planning cases, Patient 1 is drawn from the publicly available Common Optimization for Radiation Therapy (CORT) dataset \citep{Craft2015} and forms the primary analysis example throughout this study. Patients 2-8 were selected from the UKHD cohort. The CT voxel dimensions are \qtyproduct{3.0 x 3.0 x 5.0}{\mm} for Patient 1, \qtyproduct{0.97656 x 0.97656 x 3}{\mm} for Patients 2-6 and 8, and \qtyproduct{1.9531 x 1.9531 x 3}{\mm} for Patient 7. Each case contains between 19 and 26 delineated anatomical structures, including multi-level PTVs (e.g., PTV63/70) and critical OARs, such as the bilateral parotid glands and the skin.

\paragraph*{Lung}
For the lung NTCP dataset, 73 patients were retrospectively collected from a Phase II clinical trial comparing photon (IMRT) and passive-scattered proton radiotherapy (PSPT) for patients with stage IIB-IIIA non-small cell lung cancer (NSCLC; NCT00915005).\footnote{Approved by the Massachusetts General Hospital internal review board (IRB), protocol 2021P002250, \enquote{Therapy Imaging Program - Optimization of treatment outcome through retrospective data analysis}.} Patients were treated at MD Anderson Cancer Center and Massachusetts General Hospital \cite{Liao2018,Ajdari2022,Maragno2024}. Imaging data and corresponding segmentations are available for all patients; \SI{50}{\percent} of tumors were located in the right lung. Radiation pneumonitis (RP) was clinically graded according to the CTCAE v3, with grade 2 or higher (RP2+) as the primary endpoint. 
 
The seven lung planning cases were selected from this cohort. All share the same CT voxel dimensions of \qtyproduct{0.97656 x 0.97656 x 2.5}{\mm}. The number of delineated anatomical structures ranges from 19 to 51, including the GTV/PTV, lungs, heart, spinal cord, and esophagus.

\begin{table}[!htbp]
    \centering
    \caption{\textbf{Summary of the NTCP datasets}.}
    \begin{scaletable}
        \centering \renewcommand{\arraystretch}{1.3}
\renewcommand{\cellalign}{cc}
{\fontsize{11}{11} \selectfont
\begin{tabular}{l l c c c l}
    \toprule
    \thead{\textbf{Dataset}} & \thead{\textbf{Modality}} & \thead{\textbf{\# Observations}} & \thead{\textbf{\# Features}} & \thead{\textbf{Feature types}} & \thead{\textbf{Label}} \\
    \cmidrule(lr){1-1} \cmidrule(lr){2-2} \cmidrule(lr){3-3} \cmidrule(lr){4-4}
    \cmidrule(lr){5-5} \cmidrule(lr){6-6}
    Head-and-neck & photon & 153 & 6/8 & \makecell{dosiomic, radiomic} & G2+ long-term xerostomia \\
    & & & & \\[-0.8em]
    Lung & proton & \phantom{1}73 & 49/68 & \makecell{dosiomic, radiomic, \\ demographic, clinopath.} & G2+ radiopneumonitis \\
    \bottomrule
\end{tabular}}
    \end{scaletable}
    \label{tab:datasets}
\end{table}

\subsection{NTCP modeling}\label{subsec:ntcp-modeling}
All NTCP models were developed using the pyanno4rt \enquote{Amadeus} v1.0.0 modeling suite, which interfaces with scikit-learn for conventional machine learning models and TensorFlow for neural networks. Models were trained directly on the raw datasets without additional preprocessing. Bayesian hyperparameter optimization was performed via stratified $k$-fold cross-validation, with $t$ iterations aimed at maximizing the area under the curve (AUC). Model robustness and feature importance were evaluated through permutation importance~\cite{Breiman2001}, based on 30 permutations and AUC scoring. Final model performance was measured in terms of log loss, balanced accuracy, precision, recall, F1, and ROC-AUC. For both cohorts, we trained logistic regression, linear SVM, and feed-forward neural network models; for the HN cohort, we additionally trained a decision tree as a proof of concept for the hybridization approach (\Cref{subsubsec:ML-plan-comparison}).

\subsubsection{Logistic regression and linear support vector machines}
For logistic regression, the HN model ($k=3$, $t=50$) employed L1 regularization ($C\approx{66.4}$), the SAGA solver, and a tolerance of $10^{-6}$. The lung model ($k=1$, $t=1$) was tuned in a one-shot approach after extensive training failed to yield a model with practical utility (that is, a model that captures the clinical dose-response relationship), employing L2 regularization ($C\approx{40.2}$, SAGA solver, tolerance $10^{-4}$).
 
The SVM models for the HN ($k=1$, $t=1$, $C\approx 69.4$) and lung ($k=5$, $t=100$, $C\approx 4.9$) cohorts both featured linear kernels and a tolerance of $10^{-5}$.

\subsubsection{Feed-forward neural networks and decision tree}
The feed-forward neural networks were implemented with a single hidden layer, ReLU activation functions, batch normalization, and early stopping with automatic learning rate reduction. Both networks were trained for 100 epochs with a batch size of 16, using binary focal cross-entropy as the loss function. For the HN cohort ($k=5$, $t=1$), the network comprises 7 hidden neurons and \SI{10}{\percent} dropout, optimized via SGD with an initial learning rate of approximately $3.9\cdot 10^{-5}$. For the lung cohort ($k=3$, $t=50$), the network is configured with 6 hidden neurons and \SI{25}{\percent} dropout, using the Adam optimizer with an initial learning rate of about $\approx{8\cdot{10^{-6}}}$.

The decision tree model for the HN cohort ($k=3$, $t=250$) used entropy as the splitting criterion and had a maximum depth of 4.

\subsection{Planning study}\label{subsec:planning-study}
We implemented a two-step experimental workflow involving treatment plan comparison and proof-of-concept hybridization. First, inverse planning problems were independently optimized using OptiCL and pyanno4rt. By employing identical input data, ML-based NTCP models, optimization components, and variable/constraint bounds, we ensured a rigorous head-to-head comparison between both frameworks.

\subsubsection{Treatment planning setup}
All plans were optimized for 30 fractions with a bixel width / spot spacing of \SI{7}{\mm}. For the HN cases, the baseline photon plans used five equidistant gantry angles (\ang{0}, \ang{72}, \ang{144}, \ang{216}, and \ang{288}) without couch rotation, while for the lung cases, the baseline proton plans was configured with three gantry angles (\ang{45}, \ang{110}, and \ang{165}) at fixed couch angles. Proton RBE was modeled using a constant factor of 1.1. In line with the original cohort, all plans were optimized without robust optimization. This also isolates the influence of the optimization paradigm from that of uncertainty management and keeps the mixed-integer problems tractable; the implications of this choice are discussed in \Cref{subsec:disc-limitations}.

To ensure consistency across varying CT resolutions and minimize interpolation artifacts, dose calculations were performed on standardized dose grids of \qtyproduct{3x3x5}{\mm} for the HN cases and \qtyproduct{3x3x3}{\mm} for the lung cases. For visualization, the optimized three-dimensional dose distributions were interpolated to match the original treatment planning CT resolutions. \Cref{tab:geometries} summarizes the resulting geometric information relevant to optimization, including grid resolution, number of voxels, number of beamlets, and number of structures.

\begin{table}[!htbp]
    \centering
    \caption{\textbf{Patient-specific imaging and treatment planning parameters}.}

\centering \renewcommand{\arraystretch}{1.3}
\renewcommand{\cellalign}{cc}
{\fontsize{10}{10} \selectfont
\begin{threeparttable}
    \begin{tabular}{l c c c c c}
        \toprule
        \thead{\textbf{Dataset}} & \thead{\textbf{Patient}} & \thead{\makecell{\textbf{Resolution} \\ CT/Dose, in \text{[mm]}}} & \thead{\makecell{\textbf{\# Voxels} \\ CT/Dose}} & \thead{\textbf{\# Beamlets}} & \thead{\textbf{\# Structures}} \\
        \cmidrule(lr){1-1} \cmidrule(lr){2-2} \cmidrule(lr){3-3} \cmidrule(lr){4-4} \cmidrule(lr){5-5} \cmidrule(lr){6-6} 
        \multirow{8}{*}{Head-and-neck} 
        & 1 & 3.0 $\times$ 3.0 $\times$ 5.0 / 3.0 $\times$ 3.0 $\times$ 5.0 & 1,736,707 / 1,736,707 & 5,399 & 23 \\
        & 2 & 0.97656 $\times$ 0.97656 $\times$ 3.0 / 3.0 $\times$ 3.0 $\times$ 5.0 & 27,000,832 / 1,729,118 & 3,560 & 25 \\
        & 3 & 0.97656 $\times$ 0.97656 $\times$ 3.0 / 3.0 $\times$ 3.0 $\times$ 5.0 & 43,515,904 / 2,788,900 & 2,552 & 26 \\
        & 4 & 0.97656 $\times$ 0.97656 $\times$ 3.0 / 3.0 $\times$ 3.0 $\times$ 5.0 & 41,943,040 / 2,677,344 & 3,023 & 25 \\
        & 5 & 0.97656 $\times$ 0.97656 $\times$ 3.0 / 3.0 $\times$ 3.0 $\times$ 5.0 & 38,273,024 / 2,091,675 & 3,069 & 21 \\
        & 6 & 0.97656 $\times$ 0.97656 $\times$ 3.0 / 3.0 $\times$ 3.0 $\times$ 5.0 & 32,768,000 / 1,736,707 & 5,408 & 21 \\
        & 7 & 1.9531 $\times$ 1.9531 $\times$ 3.0 / 3.0 $\times$ 3.0 $\times$ 5.0 & 7,842,304 / 2,005,056 & 4,909 & 22 \\
        & 8 & 0.97656 $\times$ 0.97656 $\times$ 3.0 / 3.0 $\times$ 3.0 $\times$ 5.0 & 31,719,424 / 2,035,897 & 5,495 & 19 \\
        \midrule
        \multirow{7}{*}{Lung}
        & 1 & 0.97656 $\times$ 0.97656 $\times$ 2.5 / 3.0 $\times$ 3.0 $\times$ 3.0 & 35,651,584 / 3,151,457 & 7,496 & 31 \\
        & 2 & 0.97656 $\times$ 0.97656 $\times$ 2.5 / 3.0 $\times$ 3.0 $\times$ 3.0 & 41,943,040 / 3,709,237 & 42,391 & 27 \\
        & 3 & 0.97656 $\times$ 0.97656 $\times$ 2.5 / 3.0 $\times$ 3.0 $\times$ 3.0 & 38,273,024 / 3,374,569 & 13,895 & 51 \\
        & 4 & 0.97656 $\times$ 0.97656 $\times$ 2.5 / 3.0 $\times$ 3.0 $\times$ 3.0 & 34,078,720 / 3,012,012 & 4,794 & 22 \\
        & 5 & 0.97656 $\times$ 0.97656 $\times$ 2.5 / 3.0 $\times$ 3.0 $\times$ 3.0 & 31,457,280 / 2,788,900 & 3,941 & 40 \\
        & 6 & 0.97656 $\times$ 0.97656 $\times$ 2.5 / 3.0 $\times$ 3.0 $\times$ 3.0 & 44,040,192 / 3,904,460 & 4,285 & 51 \\
        & 7 & 0.97656 $\times$ 0.97656 $\times$ 2.5 / 3.0 $\times$ 3.0 $\times$ 3.0 & 44,040,192 / 3,904,460 & 6,231 & 19 \\
        \bottomrule
    \end{tabular}
\end{threeparttable}
}
    \label{tab:geometries}
\end{table}

\subsubsection{Objective function}
In both scenarios, the composite objective function $\mathcal{J}$ is defined as a weighted sum, $\mathcal{J}(F)=\langle w, h \rangle$, where $w$ denotes the vector of weights and $h$ the vector of individual objective function values. The baseline optimization components for conventional treatment plans, summarized in \Cref{tab:hn-components,tab:lung-components}, penalize dose excesses beyond prescribed limits through squared overdosing terms for OARs and deviations from specified reference levels through squared deviation terms for target volumes.

\begin{table}[!htbp]
    \centering
    \caption{\textbf{Optimization components for the conventional HN plan}.}
    \centering \renewcommand{\arraystretch}{1.3}
\renewcommand{\cellalign}{cc}
{\fontsize{11}{11} \selectfont
\begin{threeparttable}
    \begin{tabular}{l l c l}
        \toprule
        \thead{\textbf{Segment} \\ $\phantom{0}$} & \thead{\textbf{Component function} \\ $\mathit{f_{i}}$} & \thead{\textbf{Component weight} \\ $\mathit{w_{i}}$} & \thead{\textbf{Dose parameter} \\ $\phantom{0}$} \\
        \cmidrule(lr){1-1} \cmidrule(lr){2-2} \cmidrule(lr){3-3} \cmidrule(lr){4-4}
        Parotid (Left) & Squared Overdosing & \phantom{0}100 & \SI{25}{\gray} max. \\
        & & & \\[-0.8em]
        Parotid (Right) & Squared Overdosing & \phantom{0}100 & \SI{25}{\gray} max. \\
        & & & \\[-0.8em]
        PTV63 & Squared Deviation & 1000 & \SI{63}{\gray} ref. \\
        & & & \\[-0.8em]
        PTV70 & Squared Deviation & 1000 & \SI{70}{\gray} ref. \\
        & & & \\[-0.8em]
        Skin\tnote{*} & Squared Overdosing & \phantom{0}800 & \SI{30}{\gray} max. \\
        \bottomrule
    \end{tabular}
    \begin{tablenotes}[flushleft]
        \footnotesize
        \item[*] only considered if skin has been contoured
    \end{tablenotes}
\end{threeparttable}
}
    \label{tab:hn-components}
\end{table}

\begin{table}[!htbp]
    \centering
    \caption{\textbf{Optimization components for the conventional lung plan}.}
    \centering \renewcommand{\arraystretch}{1.3}
\renewcommand{\cellalign}{cc}
{\fontsize{11}{11} \selectfont
\begin{threeparttable}
    \begin{tabular}{l l c l}
        \toprule
        \thead{\textbf{Segment} \\ $\phantom{0}$} & \thead{\textbf{Component function} \\ $\mathit{f_{i}}$} & \thead{\textbf{Component weight} \\ $\mathit{w_{i}}$} & \thead{\textbf{Dose parameter} \\ $\phantom{0}$} \\
        \cmidrule(lr){1-1} \cmidrule(lr){2-2} \cmidrule(lr){3-3} \cmidrule(lr){4-4}
        Esophagus & Squared Overdosing & \phantom{0}100 & \SI{45}{\gray}\text{(RBE)} max. \\
        & & & \\[-0.8em]
        GTV\tnote{*} & Squared Deviation & 1000 & \SI{66}{\gray}\text{(RBE)} ref. \\
        & & & \\[-0.8em]
        Heart & Squared Overdosing & \phantom{0}100 & \SI{15.06}{\gray}\text{(RBE)} max. \\
        & & & \\[-0.8em]
        Lung\tnote{*} & Squared Overdosing & \phantom{0}100 & \SI{19.15}{\gray}\text{(RBE)} max. \\
        & & & \\[-0.8em]
        Liver\tnote{**} & Squared Overdosing & \phantom{0}100 & \SI{25}{\gray}\text{(RBE)} max. \\
        & & & \\[-0.8em]
        PTV & Squared Deviation & 1000 & \SI{65}{\gray}\text{(RBE)} ref. \\
        & & & \\[-0.8em]
        Spinal Cord & Squared Overdosing & \phantom{0}100 & \SI{54}{\gray}\text{(RBE)} max. \\
        \bottomrule
    \end{tabular}
    \begin{tablenotes}[flushleft]
        \footnotesize
        \item[*] incl. subvolumes
        \item[**] only considered if liver has been contoured
    \end{tablenotes}
\end{threeparttable}
}
    \label{tab:lung-components}
\end{table}

\subsubsection{Machine learning model-based plans and plan comparison}\label{subsubsec:ML-plan-comparison}
ML-based plans were generated by re-optimizing the baseline configurations with the NTCP models described in \Cref{subsec:ntcp-modeling}, incorporated either as objective terms in pyanno4rt or as learned constraints in OptiCL. A weight of $w=1000$ was assigned to the NTCP term, except for the HN feed-forward neural network, which received a weight of $w=800$. This choice matched the target objective weights ($w=1000$) and exceeded the OAR overdosing objective weights ($w=100$) by one order of magnitude. Preliminary runs on Patient 1 from each cohort were used to determine weights that allowed the NTCP term to measurably influence the optimization without compromising target coverage. The resulting weights were kept constant across all patients and both frameworks. Since both frameworks used identical weights, neither was favored by this choice; the sensitivity to weight selection is discussed in \Cref{subsec:disc-opt-model}. For each patient, one conventional baseline plan and three ML-based plans (logistic regression, linear SVM, and feed-forward neural network) were evaluated in both frameworks. This yields a total of 120 treatment plans (15 patients $\times$ 4 plan types $\times$ 2 frameworks). Dosimetric comparisons focused on mean dose differences for OARs and differences in $D_{95}$ (the dose received by 95\% of the volume) for the GTV/PTV, the latter serving as a measure of target coverage.

\paragraph{Hybridization}
To demonstrate MIP/NLP hybridization, we leverage the nonlinear solution of the decision tree-based HN treatment plan ($w=1000$) from pyanno4rt as a priori information for a hybrid Gurobi model within OptiCL, following the strategies described in \Cref{subsec:hybrid-strategies}. The resulting hybrid plan was compared to the standalone OptiCL solution to quantify the benefits of the hybrid approach.


\section{Results} \label{sec:results}
This section presents the results of NTCP modeling (\Cref{subsec:ntcp-modeling-results}), treatment plan comparison (\Cref{subsec:performance-comparison}), and hybridization (\Cref{subsec:hybrid-example}). Dosimetric and NTCP differences are calculated by subtracting OptiCL's from pyanno4rt's outputs, unless otherwise stated. All NTCP values reported below represent model-predicted probabilities evaluated on the optimized dose distributions. No observed clinical outcomes are available for the planning cases, nor are any used in this retrospective computational study. 

Hence, the NTCP models serve a dual purpose: they guide plan generation within each optimization framework and are subsequently applied post hoc to the resulting dose distributions for comparison. Consequently, plan evaluation relies on the same models used to guide optimization rather than on independent clinical outcome data. All post-hoc NTCP values were computed using the original model implementations (scikit-learn/Keras predictors) applied to the final dose distributions. Specifically, the decision tree example was evaluated using the original tree rather than the projection-based surrogate employed during optimization; similarly, OptiCL plans were evaluated using the original model rather than its mixed-integer encoding. Additional results appear in the Supplementary Materials.

\subsection{NTCP modeling} \label{subsec:ntcp-modeling-results}
The following paragraphs summarize the main findings; detailed results are provided in Supplementary Tables S29-S33. Notably, maximizing predictive performance was not the primary selection criterion; instead, model selection prioritized practical impact on treatment plans during the optimization process.

\paragraph*{Logistic regression and linear support vector machines}
For logistic regression (LR), the HN model achieved a training AUC of 0.89 (95\% bootstrap CI: 0.82--0.94) and a cross-validated AUC of 0.85 (0.78--0.92). Permutation importance identified the left and right parotid volumes (AUC scores: 0.120$\pm$0.049; 0.106$\pm$0.051) and the right parotid mean dose (0.081$\pm$0.043) as the three most influential features. The lung model reached a training AUC of 0.85 (0.75--0.92) but its cross-validated AUC was substantially lower at 0.37 (0.23--0.52). Its most influential features were FEV1/FVC percentages (0.103$\pm$0.037; 0.064$\pm$0.032) and the total lung mean dose (0.061$\pm$0.030).

The HN SVM model exhibited near-random predictive performance, with a training AUC of 0.49 (0.30--0.66) and a cross-validated AUC of 0.51 (0.30--0.70). It focused on left parotid sphericity (0.065$\pm$0.055) and right parotid mean dose (0.045$\pm$0.043). In contrast, the lung model achieved a training AUC of 0.88 (0.76--0.98) and a cross-validated AUC of 0.68 (0.54--0.81). The strongest predictors were right and total lung mean doses (0.013$\pm$0.004; 0.008$\pm$0.002) and chemotherapy agents (Carboplatin-Paclitaxel-Taxotere; 0.006$\pm$0.002).

\paragraph*{Feed-forward neural networks and decision tree}
For the HN cohort, the feed-forward neural network (FFNN) yielded a training AUC of 0.58 (0.38--0.76) and a cross-validated AUC of 0.39 (0.23--0.56). The most influential features identified by permutation importance included the right and left parotid mean doses (0.091$\pm$0.121; 0.000$\pm$0.001). For the lung cohort, the network reached a training AUC of 0.76 (0.64--0.87) and a cross-validated AUC of 0.46 (0.31--0.61). Top predictors were total and right lung mean doses (0.017$\pm$0.004; 0.009$\pm$0.007), followed by the Karnofsky score (0.009$\pm$0.004).

The HN decision tree model achieved a training AUC of 0.77 (0.65--0.87) and a cross-validated AUC of 0.70 (0.53--0.85). Right parotid eccentricity ranked highest in permutation importance (0.271$\pm$0.069), followed by the mean doses to the right (0.052$\pm$0.053) and left (0.036$\pm$0.051) parotid glands.

\subsection{Framework comparison} \label{subsec:performance-comparison}
Several fitted models exhibited poor or near-chance predictive performance. For the purpose of this study, they nevertheless remain valid test functions: each defines a mapping from the dose distribution (and fixed covariates) to a predicted NTCP, and both frameworks are confronted with the same mapping. The comparison therefore addresses whether global and local optimization yield consistent solutions for a given model, rather than whether the resulting plans are clinically superior. \Cref{subsec:disc-model-embeddings} discusses the clinical implications of embedding weakly predictive models in more detail. 

We evaluate the NTCP model-based plans using a two-tiered approach: an initial qualitative and semi-quantitative analysis of a representative case from each cohort---using cross-sectional CT and dose images as well as DVHs---illustrating \emph{where} the two frameworks differ in their dose distributions (spatial dose difference maps, DVHs) and thereby providing qualitative context for interpreting the aggregated metrics; and subsequently, a quantitative assessment across all eight HN and seven lung cases, determining whether these differences persist systematically across patients and plan types. The representative HN case is the publicly available Patient 1 from the CORT dataset, while the representative lung case was selected as the patient with the median number of beamlets. Target coverage is compared using differences in $D_{95}$ for the GTV/PTV. Further dosimetric analyses are provided in the Supplementary Tables S1-S28.

\subsubsection{Single-case analysis} \label{subsubsec:single-analysis}
\begin{figure}[!htbp]
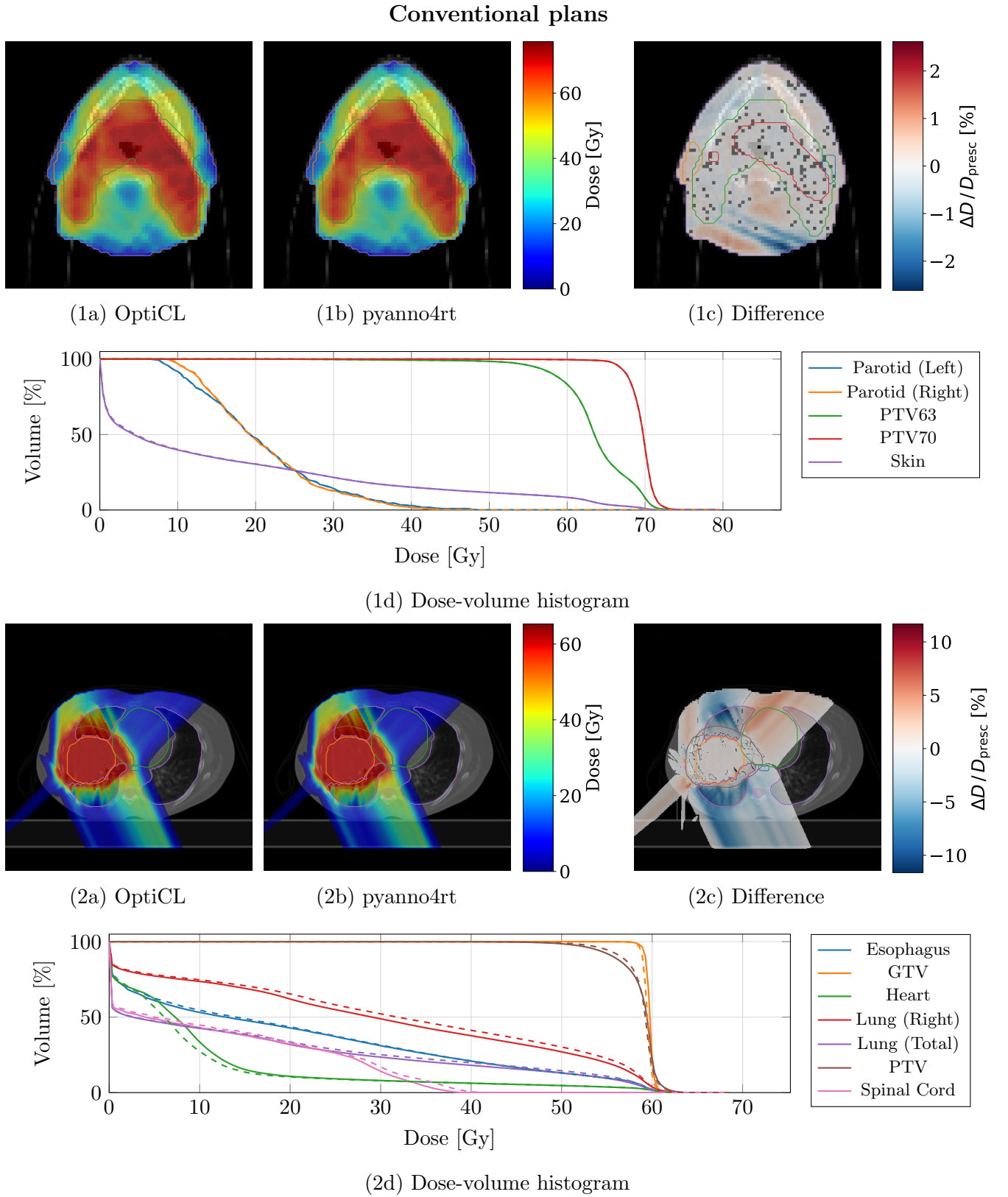

    \centering
    {\large\textbf{Conventional plans}} \\[-0.7ex]
    \input{figures/hn/hn_conv_comparison/hn_conv_comparison.tex}
    \vspace{5pt}
    \input{figures/lung/lung_conv_comparison/lung_conv_comparison.tex}
    \caption{\textbf{Comparison of conventional example plans}. For HN, dose distributions from OptiCL (1a) and pyanno4rt (1b) show negligible differences (1c), resulting in overlapping DVH curves (1d; OptiCL: \protect\solid, pyanno4rt: \protect\dashed). For lung, measurable differences arise in OAR sparing and target coverage between OptiCL (2a) and pyanno4rt (2b). A redistribution of dose between beams is observed (2c), reflected in the DVH curves (2d). In both cases, the dose difference $\Delta{D}$ is normalized to the prescribed dose $D_{\text{prescr}}$.}
    \label{fig:conv-comparison}
\end{figure}

\begin{figure}[!htbp]
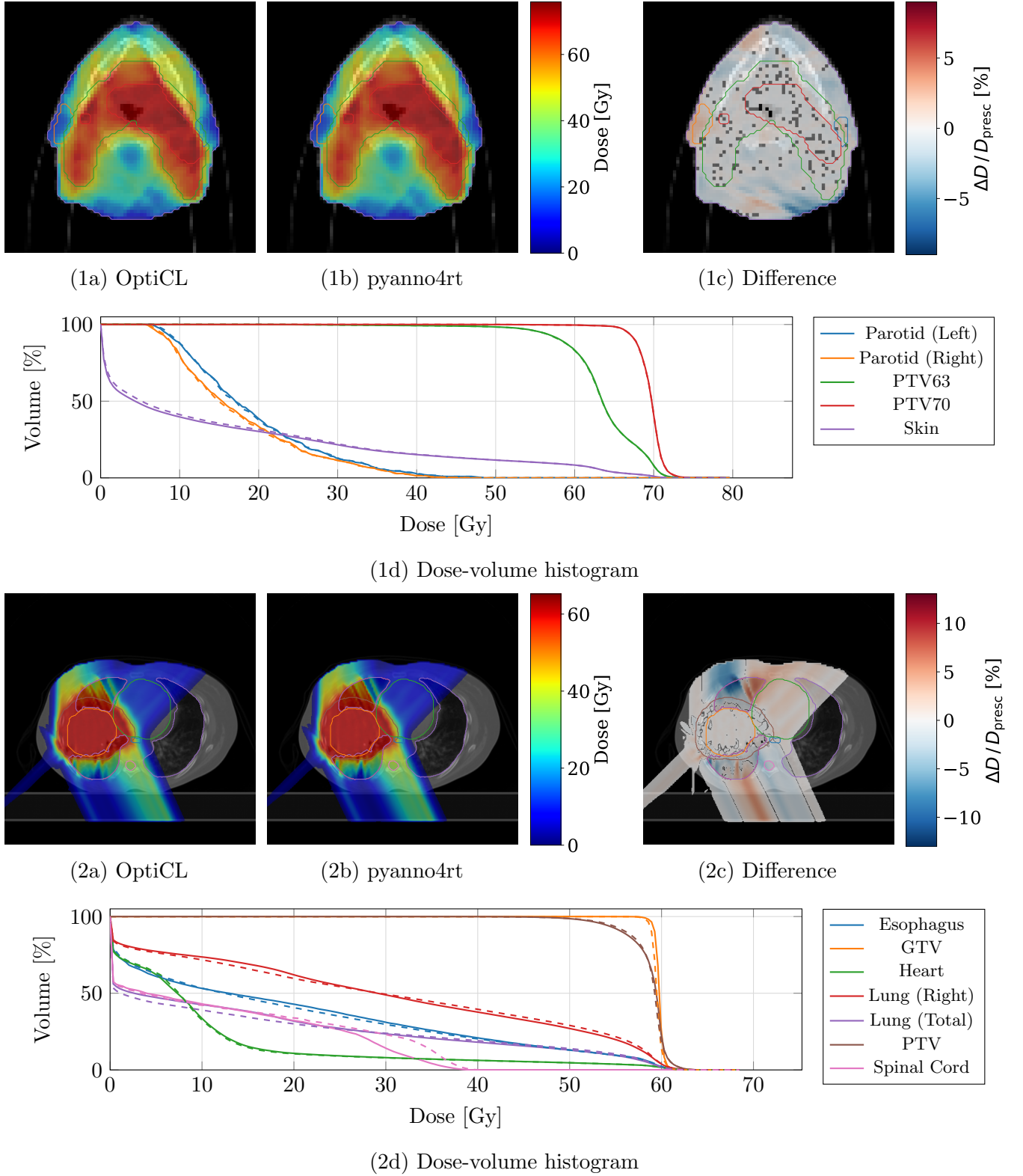

    \centering
    {\large\textbf{Logistic regression-based plans}} \\[-0.7ex]
    \input{figures/hn/hn_lr_comparison/hn_lr_comparison.tex}
    \vspace{5pt}
    \input{figures/lung/lung_lr_comparison/lung_lr_comparison.tex}
    \caption{\textbf{Comparison of logistic regression-based example plans}. For HN, dose distributions from OptiCL (1a) and pyanno4rt (1b) are highly comparable (1c). However, pyanno4rt achieves slightly lower parotid doses at the expense of an increased skin dose (1d; OptiCL: \protect\solid, pyanno4rt: \protect\dashed). For lung, comparing OptiCL (2a) and pyanno4rt (2b) shows that $D_{95}$ target coverage is maintained by a dose redistribution that prioritizes the inferior beam over the left-sided beam (2c). While most OARs receive nearly equivalent doses, OptiCL yields superior sparing of the spinal cord (2d; OptiCL: \protect\solid, pyanno4rt: \protect\dashed). In both cases, the dose difference $\Delta{D}$ is normalized to the prescribed dose $D_{\text{prescr}}$.}
     \label{fig:lr-comparison}
\end{figure}

\begin{figure}[!htbp]
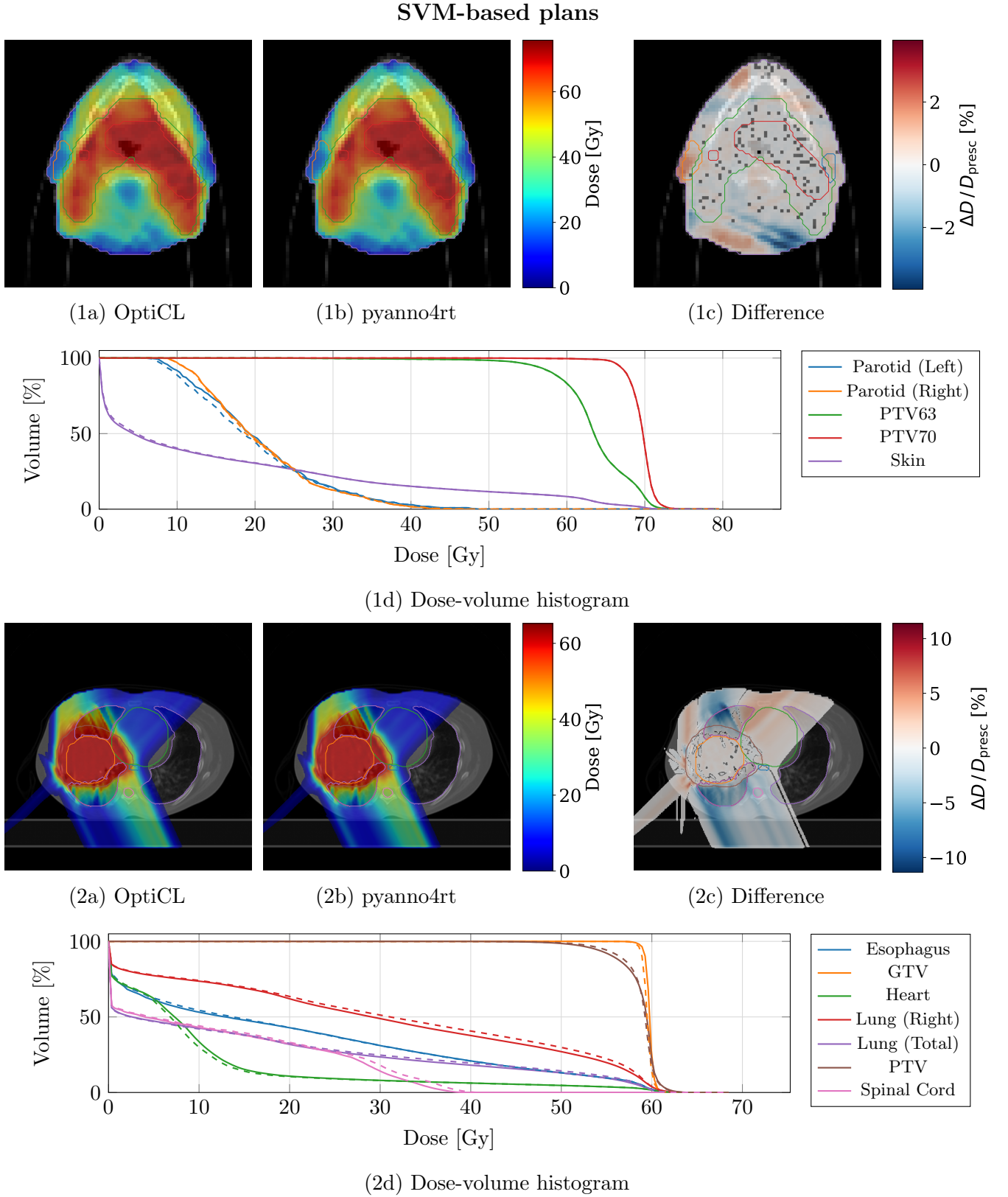

    \centering
    {\large\textbf{SVM-based plans}} \\[-0.7ex]
    \input{figures/hn/hn_svm_comparison/hn_svm_comparison.tex}
    \vspace{5pt}
    \input{figures/lung/lung_svm_comparison/lung_svm_comparison.tex}
    \caption{\textbf{Comparison of SVM-based example plans}. For the HN case (Patient 1), dose distributions from OptiCL (1a) and pyanno4rt (1b) remain highly consistent, with only minor hot and cold spots observed in the difference plot (1c). pyanno4rt achieves a reduction in the left parotid dose at the cost of a marginal increase in skin exposure (1d; OptiCL: \protect\solid, pyanno4rt: \protect\dashed). Conversely, for the lung case (Patient 1), dose differences between OptiCL (2a) and pyanno4rt (2b) resemble those of the conventional plan, as reflected in the difference plot (2c) and by the almost overlapping DVH curves (2d; OptiCL: \protect\solid, pyanno4rt: \protect\dashed). In both cases, the dose difference $\Delta{D}$ is normalized to the prescribed dose $D_{\text{prescr}}$.}
    \label{fig:svm-comparison}
\end{figure}

\begin{figure}[!htbp]
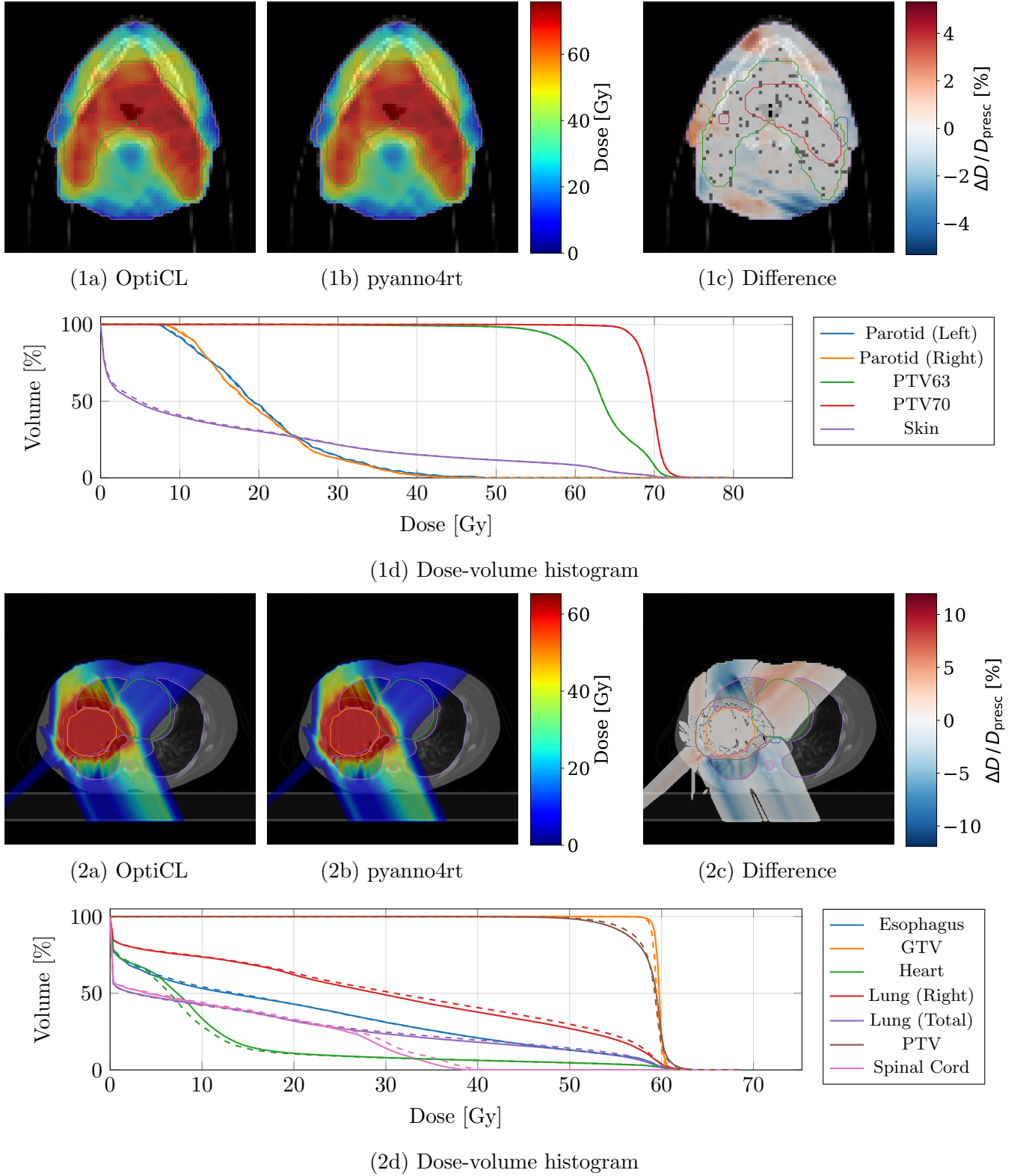

    \centering
    {\large\textbf{Neural network-based plans}} \\[-0.7ex]
    \input{figures/hn/hn_mlp_comparison/hn_mlp_comparison.tex}
    \vspace{5pt}
    \input{figures/lung/lung_mlp_comparison/lung_mlp_comparison.tex}
    \caption{\textbf{Comparison of neural network-based example plans}. For HN, dose distributions from OptiCL (1a) and pyanno4rt (1b) remain stable, with only minor differences (1c) as evidenced for skin in the DVH (1d; OptiCL: \protect\solid, pyanno4rt: \protect\dashed). For lung, OptiCL (2a) and pyanno4rt (2b) exhibit behavior consistent with the SVM-based results; OptiCL redistributes beam weights (2c) to achieve superior sparing of the lungs and spinal cord at the cost of a marginal increase in heart dose (2d; OptiCL: \protect\solid, pyanno4rt: \protect\dashed). In both cases, the dose difference $\Delta{D}$ is normalized to the prescribed dose $D_{\text{prescr}}$.}
    \label{fig:mlp-comparison}
\end{figure}

For the single-case analysis step, we examine the treatment plans generated for the representative cases in greater detail. \Cref{fig:conv-comparison,fig:lr-comparison,fig:svm-comparison,fig:mlp-comparison} show the corresponding slice images and DVHs for OptiCL and pyanno4rt.

The dose distributions for the conventional HN plans (\Cref{fig:conv-comparison}, (1a)-(1d)) are highly conformal, with near-zero mean dose differences for the left and right parotid glands (\SI{-0.01}{\gray}; \SI{-0.04}{\gray}). Target coverage, assessed by $D_{95}$, is similarly preserved, with differences of \SI{-0.02}{\gray} for PTV63 and \SI{-0.01}{\gray} for PTV70. Accordingly, the DVH curves overlap, indicating close agreement between OptiCL and pyanno4rt. For the conventional lung plans (\Cref{fig:conv-comparison}, (2a)-(2d)), differences in OAR sparing and target coverage are more pronounced: mean dose differences are \SI{-1.55}{\gray} and \SI{-0.72}{\gray} for the right and total lung, and \SI{0.66}{\gray} and \SI{-0.70}{\gray} for the heart and spinal cord. The $D_{95}$ differences for the GTV and PTV are \SI{0.32}{\gray} and \SI{-1.15}{\gray}. These deviations coincide with a redistribution of dose between beams (\Cref{fig:conv-comparison}, (2c)).

Embedding the ML models yields HN plans with closely matching dose distributions across OptiCL and pyanno4rt (\Cref{fig:lr-comparison,fig:svm-comparison,fig:mlp-comparison}, (1a)-(1d)). Mean dose differences for the left and right parotid glands are small and positive for LR (\SI{0.04}{\gray}; \SI{0.15}{\gray}) and SVM (\SI{0.66}{\gray}; \SI{0.14}{\gray}), while FFNN achieves identical mean doses in both frameworks. At the same time, the skin shows negative mean dose differences across all three models (LR: \SI{-0.48}{\gray}; SVM: \SI{-0.17}{\gray}; FFNN: \SI{-0.28}{\gray}). For the lung case, dose differences are again more pronounced: target coverage remains similar across models, amounting to \SI{0.35}{\gray}/\SI{-0.49}{\gray} for LR, \SI{0.35}{\gray}/\SI{-0.93}{\gray} for SVM, and \SI{0.32}{\gray}/\SI{-0.98}{\gray} for FFNN, respectively. These values indicate higher GTV coverage but lower PTV coverage with OptiCL. Regarding the OARs, mean dose differences in the right lung are negative across all models (LR: -\SI{0.03}{\gray}; SVM: -\SI{1.09}{\gray}; FFNN: -\SI{1.03}{\gray}), whereas those for the total lung are positive for LR (\SI{0.47}{\gray}), but negative for for SVM (\SI{-0.43}{\gray}) and FFNN (\SI{-0.40}{\gray}). Differences remain consistently negative for the spinal cord (LR: \SI{-0.98}{\gray}; SVM: \SI{-0.61}{\gray}; FFNN: \SI{-0.63}{\gray}) and positive for the heart (LR: \SI{0.03}{\gray}; SVM: \SI{0.32}{\gray}; FFNN: \SI{0.49}{\gray}). For the esophagus, the mean dose difference fluctuates (LR: \SI{0.35}{\gray}; SVM: \SI{-0.19}{\gray}; FFNN: \SI{-0.18}{\gray}).

In summary, the dose distributions generated by OptiCL and pyanno4rt show varying degrees of agreement across the two representative cases. Differences are generally smaller for the HN case, whereas more pronounced deviations are observed for the lung case, including differences in OAR sparing and target coverage. These findings characterize the agreement between the two frameworks in the selected cases.

\subsubsection{Multi-patient evaluation} \label{subsubsec:multi-analysis}
To complement the single-case analysis, we compared OptiCL and pyanno4rt across all eight HN and seven lung cases, assessing the consistency of the observed differences across patients and cohorts.

\begin{figure}[!htbp]
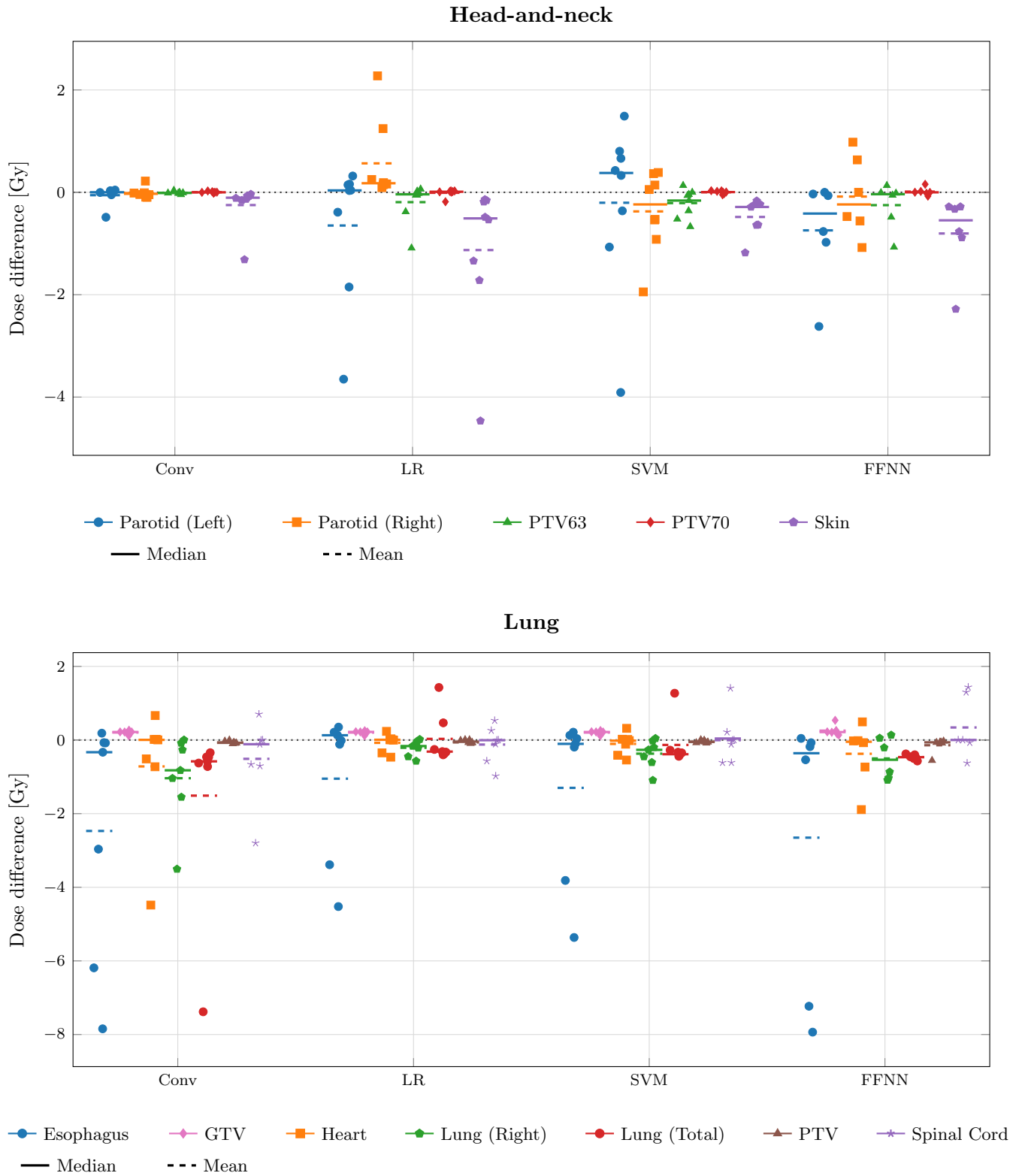

    \centering
    \input{figures/hn/HN_dose_diff}
    \vspace{15pt}
    \input{figures/lung/Lung_dose_diff}
    \caption{\textbf{Patient-wise mean dose differences by model type}. Differences are calculated as the mean total dose in OptiCL minus that in pyanno4rt for each patient. The plots display these difference for each model type (conventional, logistic regression, linear SVM, and feed-forward neural network) across all volumes of interest. In the HN cohort, differences are small for the conventional plans but increase for the parotid glands upon integrating ML models. In the lung cohort, differences are generally larger, particularly for conventional plans and for the esophagus across all model types. Most data points lie below zero, indicating lower mean total doses with OptiCL than with pyanno4rt.}
     \label{fig:dosimetrics_jitter_plots}
\end{figure}

\Cref{fig:dosimetrics_jitter_plots} (top) displays the mean dose differences stratified by patient, structure, and plan type for the HN cohort. For conventional plans, differences are close to zero across all structures (Parotid (Left): \SI{-0.06}{\gray}; Parotid (Right): \SI{-0.02}{\gray}; PTV63: \SI{-0.002}{\gray}; PTV70: \SI{-0.001}{\gray}; Skin: \SI{-0.25}{\gray}), and the differences in the target metrics (\Cref{tab:target_summary}, upper subtable) are also marginal. For ML-based plans, mean dose differences exhibit a wider spread, particularly for the left and right parotid glands, with ranges of [\SI{-3.91}{\gray}, \SI{1.49}{\gray}] and [\SI{-1.95}{\gray}, \SI{2.28}{\gray}]), respectively. Across the evaluated structures, differences vary in magnitude and direction between patients and plan types. Skin doses are consistently lower with OptiCL, whereas parotid doses are lower with either OptiCL or pyanno4rt depending on the patient and plan type. Differences in target coverage are slightly negative for PTV63 and close to zero for PTV70 across the evaluated plans, whereas the GI metric exhibits more pronounced deviations (\Cref{tab:target_summary}). 
For the lung cohort (\Cref{fig:dosimetrics_jitter_plots}, bottom),  conventional plans exhibit a wider spread of dose differences than ML-based plans. The esophagus shows pronounced outliers across plan types, with dose differences reaching \SI{-7.93}{\gray}. For the right and total lung, mean dose differences are \SI{-0.21}{\gray}/\SI{0.03}{\gray} (LR), \SI{-0.37}{\gray}/\SI{-0.13}{\gray} (SVM), and \SI{-0.50}{\gray}/\SI{-0.46}{\gray} (FFNN), respectively. In terms of target coverage, mean GTV dose differences are consistently positive (LR: \SI{0.20}{\gray}; SVM: \SI{0.20}{\gray}; FFNN: \SI{0.25}{\gray}), whereas mean PTV dose differences are slightly negative (LR: \SI{-0.06}{\gray}; SVM: \SI{-0.04}{\gray}; FFNN: \SI{-0.14}{\gray}). The corresponding D95 differences are also generally small (\Cref{tab:target_summary}, bottom subtable), with the GI metric showing the most pronounced deviations.

\begin{table}[!htbp]
    \centering
    \caption{\textbf{Paired differences (OptiCL $-$ pyanno4rt) in target coverage, homogeneity, gradient and hot-spot metrics across HN and lung patients}. Values represent the mean paired difference (OptiCL $-$ pyanno4rt) across patients, with the 95th-percentile bootstrap confidence interval reported in brackets.}

\centering \renewcommand{\arraystretch}{1.3}
\renewcommand{\cellalign}{cc}
{\fontsize{9}{9} \selectfont
\begin{threeparttable}
\begin{tabular}{l c c c c}
    \toprule
    & \multicolumn{4}{c}{\thead{\textbf{Model}}} \\

    \cmidrule(lr){2-5}

    \thead{\textbf{Metric}}
    & \thead{\textbf{Conventional}}
    & \thead{\textbf{Logistic Regression}}
    & \thead{\textbf{SVM}}
    & \thead{\textbf{Neural Network}} \\

    \cmidrule(lr){1-1}
    \cmidrule(lr){2-5}

    HN patients, $n$ & 8 & 8 & 7\textsuperscript{$\dagger$} & 6\textsuperscript{$\dagger$} \\

    PTV 63 Gy, $D_{95}$ [Gy] & -0.0 [-0.0, 0.0] & -0.3 [-0.6, -0.0] & -0.2 [-0.4, -0.0] & -0.3 [-0.6, 0.0] \\

    PTV 63 Gy, HI & 0.000 [-0.001, 0.002] & 0.007 [0.002, 0.012] & 0.010 [0.002, 0.022] & 0.016 [0.002, 0.032] \\

    PTV 70 Gy, $D_{95}$ [Gy] & 0.0 [-0.0, 0.0] & 0.0 [-0.1, 0.1] & 0.0 [-0.0, 0.0] & 0.0 [-0.0, 0.1] \\

    PTV 70 Gy, HI & 0.000 [-0.000, 0.000] & -0.001 [-0.002, -0.000] & -0.000 [-0.001, 0.000] & -0.001 [-0.003, -0.000] \\

    PTV 70 Gy, GI ($V_{50\%}/V_{100\%}$) & 0.5 [-5.3, 5.9] & 226.6 [29.0, 545.4] & 36.9 [-5.5, 102.6] & 69.3 [4.8, 167.8] \\

    $D_{\max}$ / Rx [\%] & -0.1 [-0.2, 0.0] & -2.7 [-4.9, -1.1] & 0.4 [-0.2, 1.1] & -0.2 [-1.2, 0.8] \\

    Body, $V_{>105\%}$ [cm$^3$] & -0.0 [-0.0, 0.0] & -0.2 [-0.4, -0.1] & -0.1 [-0.2, 0.0] & -0.1 [-0.4, -0.0] \\

    \bottomrule
\end{tabular}

\begin{tablenotes}[flushleft]
    \footnotesize
    \item[] $D_{95}$ in Gy; $D_{\max}$ as a percentage of the prescribed dose;
    $V_{>105\%}$ as the volume of the patient outline receiving more than 105\% of the
    prescription. HI $=(D_{2\%}-D_{98\%})/D_{50\%}$ and GI are dimensionless.
    \item[] Per-patient values are given in the supplementary tables
    (Tables~S13--S20).
    \item[\textsuperscript{$\dagger$}] The neural network model-based plan was not optimized for two
    patients, and the SVM plan of one patient did not converge and is excluded; $n$ differs
    between columns accordingly.
\end{tablenotes}
\end{threeparttable}
}

    \vspace{10pt}

\centering \renewcommand{\arraystretch}{1.3}
\renewcommand{\cellalign}{cc}
{\fontsize{9}{9} \selectfont
\begin{threeparttable}
\begin{tabular}{l c c c c}
    \toprule
    & \multicolumn{4}{c}{\thead{\textbf{Model}}} \\

    \cmidrule(lr){2-5}

    \thead{\textbf{Metric}}
    & \thead{\textbf{Conventional}}
    & \thead{\textbf{Logistic Regression}}
    & \thead{\textbf{SVM}}
    & \thead{\textbf{Neural Network}} \\

    \cmidrule(lr){1-1}
    \cmidrule(lr){2-5}

    Lung patients, $n$ & 7 & 7 & 7 & 6\textsuperscript{$\dagger$} \\

    PTV 65 Gy, $D_{95}$ [Gy] & -0.8 [-1.0, -0.7] & -0.7 [-0.8, -0.6] & -0.6 [-0.8, -0.5] & 0.0 [-0.6, 1.0] \\

    PTV 65 Gy, HI & 0.017 [0.012, 0.022] & 0.012 [0.008, 0.016] & 0.010 [0.005, 0.016] & -0.012 [-0.042, 0.009] \\

    PTV 65 Gy, GI ($V_{50\%}/V_{95\%}$) & -1150.0 [-3114.7, -78.6] & -749.3 [-2018.8, -47.2] & -587.6 [-1898.2, 162.0] & 533.0 [-44.3, 1205.5] \\

    $D_{\max}$ / Rx [\%] & 0.2 [-0.5, 1.0] & 0.5 [-0.8, 2.1] & -0.2 [-0.7, 0.3] & -2.0 [-4.1, -0.2] \\

    Body, $V_{>105\%}$ [cm$^3$] & -0.0 [-0.1, 0.0] & -0.0 [-0.1, 0.0] & -0.0 [-0.1, -0.0] & -0.0 [-0.1, -0.0] \\

    \bottomrule
\end{tabular}

\begin{tablenotes}[flushleft]
    \footnotesize
    \item[] $D_{95}$ in Gy(RBE); $D_{\max}$ as a percentage of the prescribed dose;
    $V_{>105\%}$ as the volume of the patient outline receiving more than 105\% of the
    prescription. HI $=(D_{2\%}-D_{98\%})/D_{50\%}$ and GI are dimensionless.
    \item[] GI and $D_{\max}$/Rx refer to the highest prescription in the plan
    (\SI{66}{\gray}, the GTV objective), not the \SI{65}{\gray} PTV. GI is given as
    $V_{50\%}/V_{95\%}$ rather than the usual $V_{50\%}/V_{100\%}$: these plans
    normalise below prescription, so the 100\% isodose is empty for several cases
    and the conventional ratio is undefined.
    \item[] Per-case values are given in the supplementary tables
    (Tables~S21--S28).
    \item[\textsuperscript{$\dagger$}] The neural network model-based plan was not optimized for one
    patient.
\end{tablenotes}
\end{threeparttable}
}

    \label{tab:target_summary}
\end{table}

The corresponding NTCP values for all HN and lung treatment plans are reported in \Cref{tab:outcomes}. For the HN cohort, NTCP values are highly consistent between OptiCL and pyanno4rt for LR and SVM, with OptiCL yielding lower values for SVM. The SVM and FFNN models show relatively little variation across patients, with FFNN predicting generally higher NTCP values. Two OptiCL treatment plans using FFNN could not be completed within the optimization time limit. For the lung cohort, NTCP differences between the two frameworks are more pronounced, reaching up to \SI{7.7}{\percent}. For one patient (Patient 2), the Gurobi optimizer failed to identify a feasible solution.

\begin{table}[!htbp]
    \centering
    \caption{\textbf{Predicted NTCP values across model types for the HN and lung cohorts}. Results are grouped by treatment site (HN, lung) and model type (logistic regression, linear SVM, feed-forward neural network). In the HN cohort, predicted NTCP values agree closely between OptiCL and pyanno4rt, particularly for logistic regression and linear SVM, while in the lung cohort, larger discrepancies are observed.}
    \centering \renewcommand{\arraystretch}{1.3}
\renewcommand{\cellalign}{cc}
{\fontsize{11}{11} \selectfont
\begin{threeparttable}
    \begin{tabular}{l c  c c  c c  c c}
        \toprule
        & & \multicolumn{2}{c}{\thead{\textbf{Logistic Regression}}} & \multicolumn{2}{c}{\thead{\textbf{SVM}}} & \multicolumn{2}{c}{\thead{\textbf{Neural Network}}} \\
        \cmidrule(lr){3-4} \cmidrule(lr){5-6} \cmidrule(lr){7-8}
        \thead{\textbf{Dataset}} & \thead{\textbf{Patient}} & \textit{OptiCL} & \textit{pyanno4rt} & \textit{OptiCL} & \textit{pyanno4rt} & \textit{OptiCL} & \textit{pyanno4rt} \\
        \cmidrule(lr){1-1} \cmidrule(lr){2-2} \cmidrule(lr){3-4} \cmidrule(lr){5-6} \cmidrule(lr){7-8}
        \multirow{8}{*}{\makecell{Head-and-neck}} 
        & 1 & 6.9 & 6.8 & 4.9 & 5.0 & 42.7 & 46.3 \\ 
        & 2 & 13.9 & 13.7 & 4.8 & 5.0 & 47.7 & 46.4 \\ 
        & 3 & 30.7 & 30.4 & 5.5 & 5.8 & 47.7 & 48.0 \\ 
        & 4 & 1.2 & 1.2 & 5.8 & 5.9 & 48.0 & 47.3 \\ 
        & 5 & 4.1 & 4.2 & 5.7 & 5.9 & --\tnote{*} & 50.0 \\ 
        & 6 & 17.1 & 16.9 & 5.6 & 5.7 & --\tnote{*} & 50.1 \\ 
        & 7 & 16.1 & 16.3 & --\tnote{*} & 5.5 & 46.1 & 46.1 \\ 
        & 8 & 40.3 & 39.9 & 5.6 & 5.9 & 46.9 & 46.3 \\ 
        \midrule
        \multirow{7}{*}{Lung}
        & 1 & 29.3 & 33.3 & 23.0 & 25.0 & 35.0 & 42.7 \\ 
        & 2 & 5.1 & 5.2 & 13.8 & 14.2 & --\tnote{*} & 44.9 \\ 
        & 3 & 3.0 & 3.9 & 19.6 & 20.5 & 26.1 & 28.4 \\ 
        & 4 & 5.6 & 7.5 & 28.8 & 30.5 & 48.9 & 49.1 \\ 
        & 5 & 4.8 & 6.2 & 28.3 & 29.7 & 42.4 & 48.5 \\ 
        & 6 & 1.8 & 2.1 & 11.6 & 11.9 & 19.6 & 20.3 \\ 
        & 7 & 0.7 & 0.8 & 9.5 & 9.8 & 30.3 & 32.1 \\ 
        \bottomrule
    \end{tabular}
    \begin{tablenotes}[flushleft]
        \footnotesize
        \item[*] no feasible solution found / timed out
    \end{tablenotes}
\end{threeparttable}
}
    \label{tab:outcomes}
\end{table}

Overall, OptiCL achieves comparable or greater dosimetric improvements and NTCP reductions than pyanno4rt, at the cost of substantially longer optimization runtimes, typically by approximately one or two orders of magnitude. \Cref{tab:runtimes} compares the computational efficiency of the two frameworks based on wall-clock runtimes (upper subtable) and the corresponding numbers of iterations (lower subtable).

\begin{table}[!htbp]
    \centering
    \caption{\textbf{Wall-clock runtimes and number of iterations across model types for the HN and lung cohorts.} Results are grouped by treatment site (HN, lung) and model type (logistic regression, linear SVM, feed-forward neural network). pyanno4rt generally exhibits shorter runtimes than OptiCL, except for HN Patient 7, but usually requires a larger number of iterations. Runtimes remain broadly consistent across patients when logistic regression or linear SVM models are incorporated. In contrast, neural network integration substantially increases computational time in OptiCL due to the introduction of binary variables, whereas runtimes in pyanno4rt remain comparatively stable.}
    \centering \renewcommand{\arraystretch}{1.3}
\renewcommand{\cellalign}{cc}
{\fontsize{10}{10} \selectfont
\begin{threeparttable}
    \begin{tabular}{l c  c c  c c  c c  c c}
        \toprule
        & & \multicolumn{2}{c}{\thead{\textbf{Conventional}}} & \multicolumn{2}{c}{\thead{\textbf{Logistic Regression}}} & \multicolumn{2}{c}{\thead{\textbf{SVM}}} & \multicolumn{2}{c}{\thead{\textbf{Neural Network}}} \\
        \cmidrule(lr){3-4} \cmidrule(lr){5-6} \cmidrule(lr){7-8} \cmidrule(lr){9-10}
        \thead{\textbf{Dataset}} & \thead{\textbf{Patient}} & \textit{OptiCL} & \textit{pyanno4rt} & \textit{OptiCL} & \textit{pyanno4rt} & \textit{OptiCL} & \textit{pyanno4rt} & \textit{OptiCL} & \textit{pyanno4rt} \\
        \cmidrule(lr){1-1} \cmidrule(lr){2-2} \cmidrule(lr){3-4} \cmidrule(lr){5-6} \cmidrule(lr){7-8} \cmidrule(lr){9-10}
        \multirow{10}{*}{\makecell{Head-and-neck}} 
        & 1 & 1,245  & 78  & 1,630 & 50 & 1,221 & 52 & 1,558 & 35 \\
        & 2 & 422 & 68 & 469 & 46 & 432 & 76 & 18,788 & 59 \\ 
        & 3 & 236 & 55 & 273 & 61 & 256 & 42 & 16,591 & 44 \\
        & 4 & 356 & 52 & 408 & 66 & 383 & 43 & 27,925 & 45 \\
        & 5 & 362 & 79 & 385 & 48 & 388 & 46 & --\tnote{*} & 35 \\ 
        & 6 & 431 & 87 & 485 & 51 & 416 & 53 & --\tnote{*} & 42 \\ 
        & 7 & 183 & 91 & 112 & 91 & --\tnote{*} & 104 & 533 & 101 \\ 
        & 8 & 1,211 & 132 & 2,053 & 163 & 1,150 & 118 & 63,014 & 104 \\ 
        \midrule
        \multirow{7}{*}{Lung}
        & 1 & 787 & 162 & 869 & 267 & 868 & 137 & 20,059 & 190 \\ 
        & 2 & 4,838 & 1,276 & 11,023 & 1,460 & 10,513 & 475 & --\tnote{*} & 976 \\ 
        & 3 & 171 & 74 & 194 & 86 & 195 & 68 & 11,335 & 50 \\ 
        & 4 & 136 & 130 & 142 & 117 & 135 & 46 & 10,583 & 54 \\ 
        & 5 & 133 & 124 & 138 & 115 & 146 & 73 & 2,745 & 23 \\ 
        & 6 & 131 & 110 & 137 & 106 & 131 & 68 & 6,783 & 68 \\ 
        & 7 & 296 & 202 & 326 & 267 & 306 & 133 & 9,930 & 42 \\ 
        \bottomrule
    \end{tabular}
    \vspace{5pt}
    \begin{tabular}{l c  c c  c c  c c  c c}
        \toprule
        & & \multicolumn{2}{c}{\thead{\textbf{Conventional}}} & \multicolumn{2}{c}{\thead{\textbf{Logistic Regression}}} & \multicolumn{2}{c}{\thead{\textbf{SVM}}} & \multicolumn{2}{c}{\thead{\textbf{Neural Network}}} \\
        \cmidrule(lr){3-4} \cmidrule(lr){5-6} \cmidrule(lr){7-8} \cmidrule(lr){9-10}
        \thead{\textbf{Dataset}} & \thead{\textbf{Patient}} & \textit{OptiCL} & \textit{pyanno4rt} & \textit{OptiCL} & \textit{pyanno4rt} & \textit{OptiCL} & \textit{pyanno4rt} & \textit{OptiCL} & \textit{pyanno4rt} \\
        \cmidrule(lr){1-1} \cmidrule(lr){2-2} \cmidrule(lr){3-4} \cmidrule(lr){5-6} \cmidrule(lr){7-8} \cmidrule(lr){9-10}
        \multirow{10}{*}{\makecell{Head-and-neck}} 
        & 1 & 41  & 119  & 51 & 51 & 39 & 84 & 43 & 55 \\ 
        & 2 & 45 & 129 & 51 & 28 & 43 & 138 & 46 & 68 \\ 
        & 3 & 44 & 137 & 51 & 93 & 42 & 70 & 45 & 42 \\ 
        & 4 & 44 & 99 & 52 & 83 & 44 & 67 & 46 & 34 \\ 
        & 5 & 46 & 166 & 49 & 41 & 44 & 48 & --\tnote{*} & 24 \\ 
        & 6 & 46 & 156 & 55 & 47 & 46 & 87 & --\tnote{*} & 21 \\ 
        & 7 & 30 & 204 & 36 & 63 & --\tnote{*} & 64 & 32 & 57 \\ 
        & 8 & 49 & 152 & 61 & 147 & 48 & 104 & 53 & 99 \\ 
        \midrule
        \multirow{7}{*}{Lung}
        & 1 & 42 & 274 & 55 & 368 & 56 & 216 & 50 & 255 \\ 
        & 2 & 60 & 491 & 57 & 479 & 55 & 144 & --\tnote{*} & 312 \\ 
        & 3 & 37 & 212 & 56 & 189 & 54 & 155 & 42 & 107 \\ 
        & 4 & 35 & 327 & 50 & 240 & 47 & 103 & 40 & 116 \\ 
        & 5 & 38 & 387 & 50 & 305 & 52 & 226 & 44 & 28 \\ 
        & 6 & 32 & 307 & 43 & 250 & 42 & 167 & 39 & 160 \\ 
        & 7 & 38 & 402 & 53 & 429 & 47 & 223 & 46 & 72 \\ 
        \bottomrule
    \end{tabular}
    \begin{tablenotes}[flushleft]
        \footnotesize
        \item[*] no feasible solution found / timed out
    \end{tablenotes}
\end{threeparttable}
}
    \label{tab:runtimes}
\end{table}

\subsection{A hybrid decision tree-based example} \label{subsec:hybrid-example}
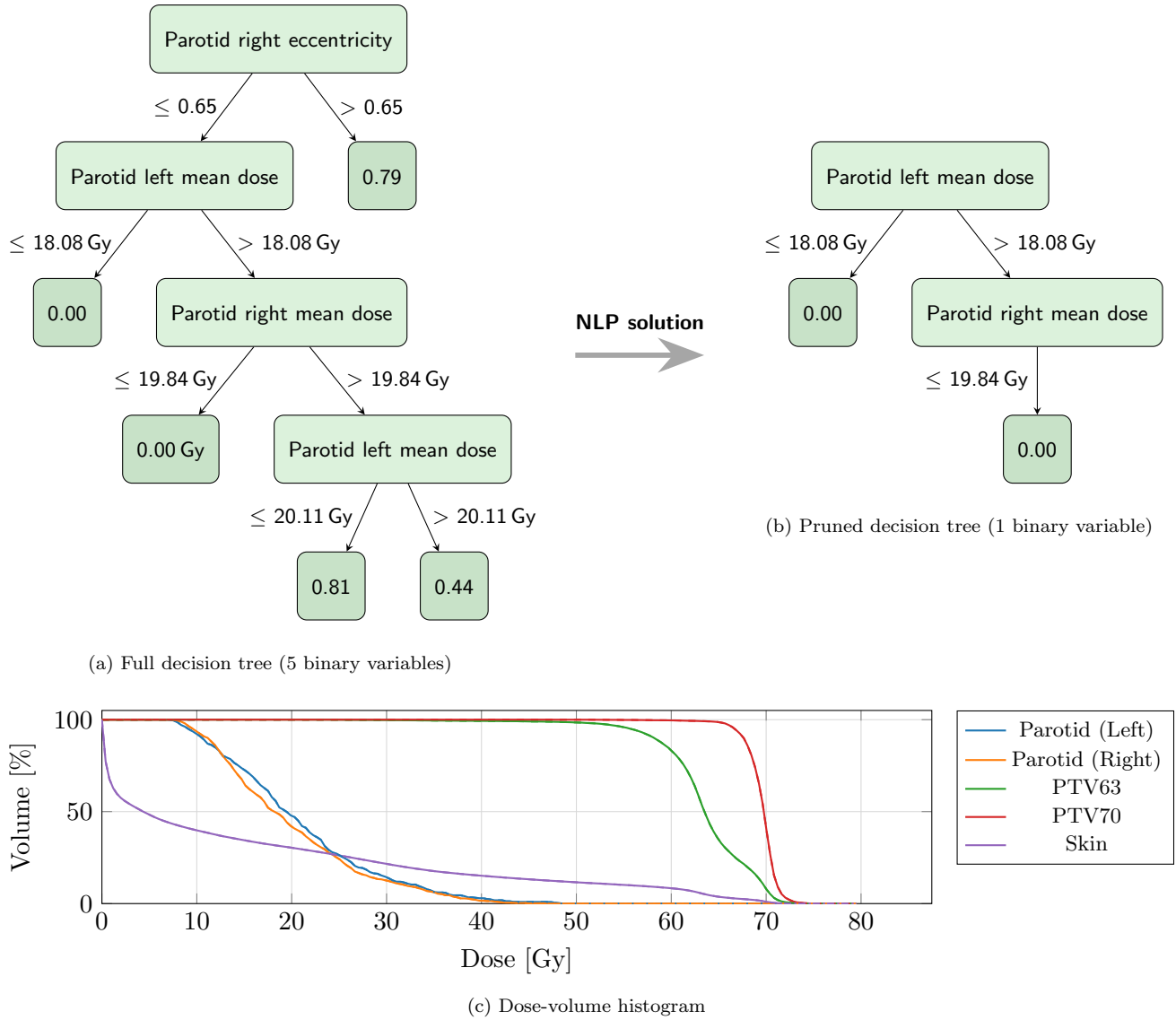
\begin{figure}[!htbp]
    \centering
    {\large\textbf{Decision tree hybridization}} \\[2ex]
    \renewcommand{\arraystretch}{1.5}
\begin{tabular}{ccc}
\centering
\begin{minipage}[c]{0.45\textwidth}
\centering
\begin{forest} tree_style
  [Parotid right eccentricity
    [Parotid left mean dose, edge label={node[midway, left, font=\footnotesize\sffamily]{$\leq$ \SI{0.65}{}}}
      [\SI{0.00}{}, edge label={node[midway, left, font=\footnotesize\sffamily]{$\leq$ \SI{18.08}{\gray}}}]
      [Parotid right mean dose, edge label={node[midway, right, font=\footnotesize\sffamily]{$>$ \SI{18.08}{\gray}}}
        [\SI{0.00}{\gray}, edge label={node[midway, left, font=\footnotesize\sffamily]{$\leq$ \SI{19.84}{\gray}}}]
        [Parotid left mean dose, edge label={node[midway, right, font=\footnotesize\sffamily]{$>$ \SI{19.84}{\gray}}}
          [\SI{0.81}{}, edge label={node[midway, left, font=\footnotesize\sffamily]{$\leq$ \SI{20.11}{\gray}}}]
          [\SI{0.44}{}, edge label={node[midway, right, font=\footnotesize\sffamily]{$>$ \SI{20.11}{\gray}}}]
        ]
      ]
    ]
    [\SI{0.79}{}, edge label={node[midway, right, font=\footnotesize\sffamily]{$>$ \SI{.65}{}}}]
  ]
\end{forest} \\[1em]
\scriptsize{(a) Full decision tree (5 binary variables)}
\end{minipage}

& 
\begin{minipage}[c]{0.08\textwidth}
\centering
\begin{tikzpicture}
    \draw[-{Stealth[scale=1.5]}, line width=2.2pt, gray!70] (0,0) -- (1.9,0) node[midway, above=6pt, font=\sffamily\footnotesize\bfseries, text=black] {NLP solution};
\end{tikzpicture}
\end{minipage}

& 
\begin{minipage}[c]{0.45\textwidth}
\centering
\begin{forest} tree_style
    [Parotid left mean dose, edge label={node[midway, left, font=\footnotesize\sffamily]{$\leq$ 0.65}}
      [\SI{0.00}{}, edge label={node[midway, left, font=\footnotesize\sffamily]{$\leq$ \SI{18.08}{\gray}}}]
      [Parotid right mean dose, edge label={node[midway, right, font=\footnotesize\sffamily]{$>$ \SI{18.08}{\gray}}}
        [\SI{0.00}{}, edge label={node[midway, left, font=\footnotesize\sffamily]{$\leq$ \SI{19.84}{\gray}}}]
      ]
    ]
\end{forest} \\[1em]
\scriptsize{(b) Pruned decision tree (1 binary variable)}
\end{minipage}
\vspace{10pt}
\end{tabular}
%
%
%

  \definecolor{clrPAROTIDLT}{rgb}{0.122,0.467,0.706}
  \definecolor{clrPAROTIDRT}{rgb}{1.000,0.498,0.055}
  \definecolor{clrPTV63}{rgb}{0.173,0.627,0.173}
  \definecolor{clrPTV70}{rgb}{0.839,0.153,0.157}
  \definecolor{clrSKIN}{rgb}{0.580,0.404,0.741}

  \begin{minipage}[t]{\textwidth}
    \centering
    \begin{tikzpicture}
      \begin{axis}[
        width=0.78\textwidth,
        height=0.25\textwidth,
        xlabel={Dose [Gy]},
        ylabel={Volume [\%]},
        xmin=0, ymin=0, ymax=105,
        grid=major,
        grid style={gray!30},
        legend pos=outer north east,
        legend style={font=\footnotesize},
      ]

      \addplot[clrPAROTIDLT, thick] table[x=dose_gy, y=vol_a_pct, col sep=comma]{figures/hn/hn_cart_comparison/dvh/dvh_PAROTID_LT.csv};
      \addlegendentry{Parotid (Left)}
      \addplot[clrPAROTIDRT, thick] table[x=dose_gy, y=vol_a_pct, col sep=comma]{figures/hn/hn_cart_comparison/dvh/dvh_PAROTID_RT.csv};
      \addlegendentry{Parotid (Right)}
      \addplot[clrPTV63, thick] table[x=dose_gy, y=vol_a_pct, col sep=comma]{figures/hn/hn_cart_comparison/dvh/dvh_PTV63.csv};
      \addlegendentry{PTV63}
      \addplot[clrPTV70, thick] table[x=dose_gy, y=vol_a_pct, col sep=comma]{figures/hn/hn_cart_comparison/dvh/dvh_PTV70.csv};
      \addlegendentry{PTV70}
      \addplot[clrSKIN, thick] table[x=dose_gy, y=vol_a_pct, col sep=comma]{figures/hn/hn_cart_comparison/dvh/dvh_SKIN.csv};
      \addlegendentry{Skin}

      \addplot[clrPAROTIDLT, thick, dashed, forget plot] table[x=dose_gy, y=vol_b_pct, col sep=comma]{figures/hn/hn_cart_comparison/dvh/dvh_PAROTID_LT.csv};
      \addplot[clrPAROTIDRT, thick, dashed, forget plot] table[x=dose_gy, y=vol_b_pct, col sep=comma]{figures/hn/hn_cart_comparison/dvh/dvh_PAROTID_RT.csv};
      \addplot[clrPTV63, thick, dashed, forget plot] table[x=dose_gy, y=vol_b_pct, col sep=comma]{figures/hn/hn_cart_comparison/dvh/dvh_PTV63.csv};
      \addplot[clrPTV70, thick, dashed, forget plot] table[x=dose_gy, y=vol_b_pct, col sep=comma]{figures/hn/hn_cart_comparison/dvh/dvh_PTV70.csv};
      \addplot[clrSKIN, thick, dashed, forget plot] table[x=dose_gy, y=vol_b_pct, col sep=comma]{figures/hn/hn_cart_comparison/dvh/dvh_SKIN.csv};


      \end{axis}
    \end{tikzpicture}
    \par\smallskip\scriptsize (c) Dose-volume histogram
  \end{minipage}

    \caption{\textbf{Decision tree hybridization}. The full decision tree fitted to the HN dataset (a) contains five leafs and branches across three features: right parotid eccentricity and the mean doses to the left and right parotid glands. By injecting the NLP solution of \SI{0.0}{\percent} to inform the Gurobi solver that a zero-NTCP state is attainable, the tree can be effectively pruned by removing nonzero leafs (b). This reduces the number of binary variables by \SI{80}{\percent} (from five to one) without compromising the final dosimetric results (c).}
    \label{fig:hybrid-decision-tree}
\end{figure}

To validate the proposed hybridization strategies, we optimized a decision tree model-based HN plan in OptiCL under two configurations: standalone optimization and hybrid optimization informed by pyanno4rt's NLP solution. By leveraging the optimal fluence candidate and predicted NTCP from pyanno4rt, we applied primal warm-start, constraint strengthening, and pruning, as described in \Cref{subsec:hybrid-strategies}. These strategies reduced the number of binary variables required to represent the decision tree by \SI{80}{\percent}, leaving only a single binary variable in the resulting MIP formulation. As shown by the DVH curves in \Cref{fig:hybrid-decision-tree}, both configurations yield equivalent dosimetric results. The wall-clock runtimes were 11078\,s for standalone optimization and 12619\,s for hybrid optimization, while the Gurobi-reported MIP gap decreased from \SI{0.02}{\percent} to \SI{0.00}{\percent}. \Cref{fig:hybrid-decision-tree} visualizes the corresponding decision tree structures.

\newpage
\section{Discussion} \label{sec:discussion}
This study compares two complementary strategies for embedding ML-based NTCP models into inverse planning: global mixed-integer optimization, implemented via OptiCL, and nonlinear local optimization, utilizing pyanno4rt. Both tools are publicly available as open-source frameworks.

Across 120 treatment plans, both frameworks produced broadly consistent dose distributions and clinically plausible trade-offs. OptiCL more frequently translated embedded ML objectives into modestly improved OAR sparing and lower predicted NTCP. Conversely, pyanno4rt achieved comparable results at a fraction of the computational cost. From a methodological standpoint, these findings suggest that the central challenge is not the displacement of one paradigm by the other, but rather their hybridization to synergistically exploit their respective strengths.

\subsection{A complementary perspective on machine learning-guided planning} \label{subsec:disc-compl-paradigms}
The comparison between OptiCL and pyanno4rt revealed close agreement for conventional plans, with increasing divergence upon the introduction of ML-based---and thus nonconvex---components. In the HN cohort, these differences remained small on average, primarily affecting OAR sparing rather than target coverage. On the other hand, deviations in the the lung cohort were more pronounced; here, OptiCL more frequently achieved lower lung and esophagus doses and, in turn, lower predicted pneumonitis risks. Collectively, these results confirm the technical robustness of both frameworks while highlighting their distinct responses to the nonconvexity inherent in ML-integrated optimization.

These differences are expected from the underlying optimization paradigms. MIP explores the feasible region using linear and mixed-integer reformulations, enabling it---in principle---to converge toward globally optimal solutions within its model representation. In contrast, NLP leverages efficient gradient-based local optimization, making it inherently sensitive to initialization and local geometry. The observed runtime gap of approximately one to two orders of magnitude is the practical manifestation of this algorithmic distinction. Consequently, NLP is the more natural choice for clinical workflows requiring rapid plan generation, repeated replanning, or interactive weight adjustments. MIP, however, remains the preferred approach when rigorous optimality guarantees are desired, such as for offline benchmarking, the verification of approximate solvers (for instance, the solvers within pyanno4rt), or the systematic study of the behavior of particular ML model classes under exact reformulations.

Nonetheless, the hybridization results suggest that the choice between these paradigms need not be an either-or decision. In the decision tree example, primal warm-starts, constraint strengthening, and pruning based on the NLP solution substantially simplified the MIP formulation. However, whether a tighter integration of both paradigms within a single algorithm would yield consistent performance gains remains an open question. Modern MIP solvers already implement sophisticated primal heuristics and search strategies, including domain filtering, which can tighten variable bounds, fix binary variables, and eliminate redundant constraints. Such techniques may be particularly effective for neural networks, where analyzing the lower and upper bounds of pre-activation values can help identify inactive or stable neurons and thereby reduce the search space. More broadly, hybrid strategies may be most effective when tailored to the architecture and complexity of the underlying ML model and applied individually or in combination to progressively restrict the MIP search space. Beyond the warm-starting, strengthening, and pruning techniques considered in this work, other established MIP heuristics may offer promising avenues for further research. These include the \emph{Relaxation Induced Neighborhood Search (RINS)} \cite{Danna2005}, which exploits agreement between an integer-feasible solution and the corresponding LP relaxation to define a neighborhood search, and the \emph{Feasibility Pump} \cite{Fischetti2005}, which iteratively alternates between rounding and continuous optimization to seek feasible integer solutions.

\subsection{Clinical trade-offs} \label{subsec:disc-clinical-tradeoffs}
From a clinical perspective, the most salient finding is that while dosimetric differences between both frameworks were generally small, the corresponding NTCP differences could still be non-negligible, especially in the lung cohort. In the HN cases, mean differences in parotid dose and predicted xerostomia remained modest, suggesting that both frameworks converge to very similar trade-offs within the more tightly constrained HN geometry. For the lung cohort, however, OptiCL achieved lower predicted pneumonitis risks in several cases at the cost of a marginal reduction in peripheral target coverage. The clinical acceptability of such a trade-off depends upon the specific biological endpoint, the magnitude of the coverage loss, and the predictive reliability of the underlying NTCP model.

In this context, the observed reductions in PTV-related dose metrics were small in absolute terms, whereas the differences in predicted lung NTCP occasionally reached levels that would be considered clinically relevant in model-based treatment selection frameworks \cite{Langendijk2021}. However, this study does not claim that lower model-predicted NTCP values inherently translate into superior clinical plans---especially given the moderate predictive performance of certain embedded models. Instead, it demonstrates that global and local optimization can induce measurably different biological trade-offs, even when the underlying physical dose distributions appear broadly consistent.

The larger framework differences in lung cases compared to HN cases are plausibly explained by anatomy, radiation modality and model architecture. Lung plans, for instance, offer greater degrees of freedom for dose redistribution, particularly given the lower integral dose inherent to proton therapy. Finally, the lung NTCP models in this study utilized a substantially larger feature set than the HN models; this increased dimensionality of the resulting optimization landscape potentially enhances the prevalence of local minima.

\subsection{Optimization model} \label{subsec:disc-opt-model}
The optimization model from \Cref{eq:optimization-model} was formulated using a weighted sum of dosimetric and model-based terms. While well-established in treatment planning and easy to implement, this approach is known to be highly dependent on the selected weights and generally fails to recover unsupported regions of the Pareto front \cite{Craft2006}. More sophisticated multi-criteria optimization strategies---such as Pareto navigation, lexicographic formulations, or $\varepsilon$-constraint methods---could provide a richer view of the trade-offs between target coverage, OAR sparing, and predicted complication risk \cite{Bokrantz2013,Kierkels2014,vanHaveren2017,Breedveld2019,Cao2022,Schubert2026}. In the present study, however, we deliberately chose a weighted-sum formulation to effectively isolate differences attributable to the optimization paradigm from potential confounding variables. This simplification is necessary to establish baseline differences between MIP and NLP embeddings, but future research should explore whether our findings generalize to more advanced optimization strategies.

\subsection{NTCP model embeddings} \label{subsec:disc-model-embeddings}
A key finding of this work is that the mathematical complexity of ML-based inverse planning is not solely determined by the prediction model, but by the interplay between the model architecture and the dose-derived input features. Even structurally simple models which yield convex (linear) prediction functions, such as logistic regression or linear SVMs, can induce nonconvex optimization landscapes when coupled with nonconvex input quantities like DVH metrics, gEUD-type measures, or shape descriptors (cf., \cite{Gabrys2018}). Conversely, highly expressive models for which global convexity with respect to $x$ is not an inherent property may remain computationally tractable if paired with structured or approximately convex feature mappings. In the case of neural networks, for example, input convexity of the prediction function can only be guaranteed under stringent structural constraints---such as the nonnegativity of the weights $W_{l}$ for $l>1$---and specific monotonicity requirements for the activation functions $a_{l}$ \cite{Amos2017,Siva2021}.

In this context, the projection-based gradient approximation for decision trees serves as a valuable complement to exact MIP tree encodings. Its primary advantage lies in preserving the original tree semantics without necessitating a smoothing of the discrete split logic. Unlike, for example, soft decision trees \cite{Irsoy2012}, which require fitting a separate differentiable model that may not maintain a strict one-to-one correspondence with the original hard tree, our approach directly utilizes the existing model. Furthermore, it provides a geometrically interpretable descent direction for gradient-based optimization.

This interpretation is further reinforced by the hybrid decision tree experiment. In this case, the NLP solution provided sufficiently informative a priori guidance to eliminate 80\% of the binary decision variables in the subsequent MIP model without degrading plan quality. This result is particularly encouraging, as it suggests that local optimization can recover structurally relevant information even in the absence of global optimality guarantees. More generally, it points toward a synergistic workflow where gradient-based optimization identifies an informative neighborhood, while exact mixed-integer optimization is reserved for the refined and more tractable subproblem.

A second aspect that merits careful discussion is the predictive performance of the embedded ML models. Three of the seven embedded models performed at or below chance level on the test data (HN SVM: AUC 0.51; lung FFNN: 0.48; lung LR: 0.37), and the HN FFNN only marginally above (0.53). We deliberately did not exclude these models, since the primary aim was to compare how two optimization paradigms treat a given model, and every fitted model is a well-defined test function for this purpose. Clinically, however, plans optimized on such models clearly lack the benefit of trustworthiness: a lower predicted NTCP under a non-predictive model carries no information on the true complication risk, and the dose redistributions induced by the model are, in effect, arbitrary with respect to the endpoint, even though they remain dosimetrically acceptable in our experiments because the conventional target and OAR objectives were retained. We therefore stress that (i) none of the plans in this study should be regarded as clinically superior on the basis of predicted NTCP, (ii) embedded models require rigorous external validation, calibration and, ideally, uncertainty quantification before clinical use, and (iii) conventional dosimetric objectives and constraints should always be retained alongside the model term as a safeguard against optimizer exploitation. Specifically, in light of the limited test set accuracy exhibited by some models in the present benchmark, the risk of optimizer exploitation becomes a critical concern if applied beyond a methodological comparison. In the absence of established dosimetric constraints, the optimizer may steer the solution toward regions where the model's generalization is unreliable, potentially yielding artificially favourable NTCP values based on inaccurate extrapolations. 

\begin{figure}[!htbp]
    \centering
    {\large\textbf{\enquote{Bad} logistic regression-based plans}} \\[-0.7ex]
    \input{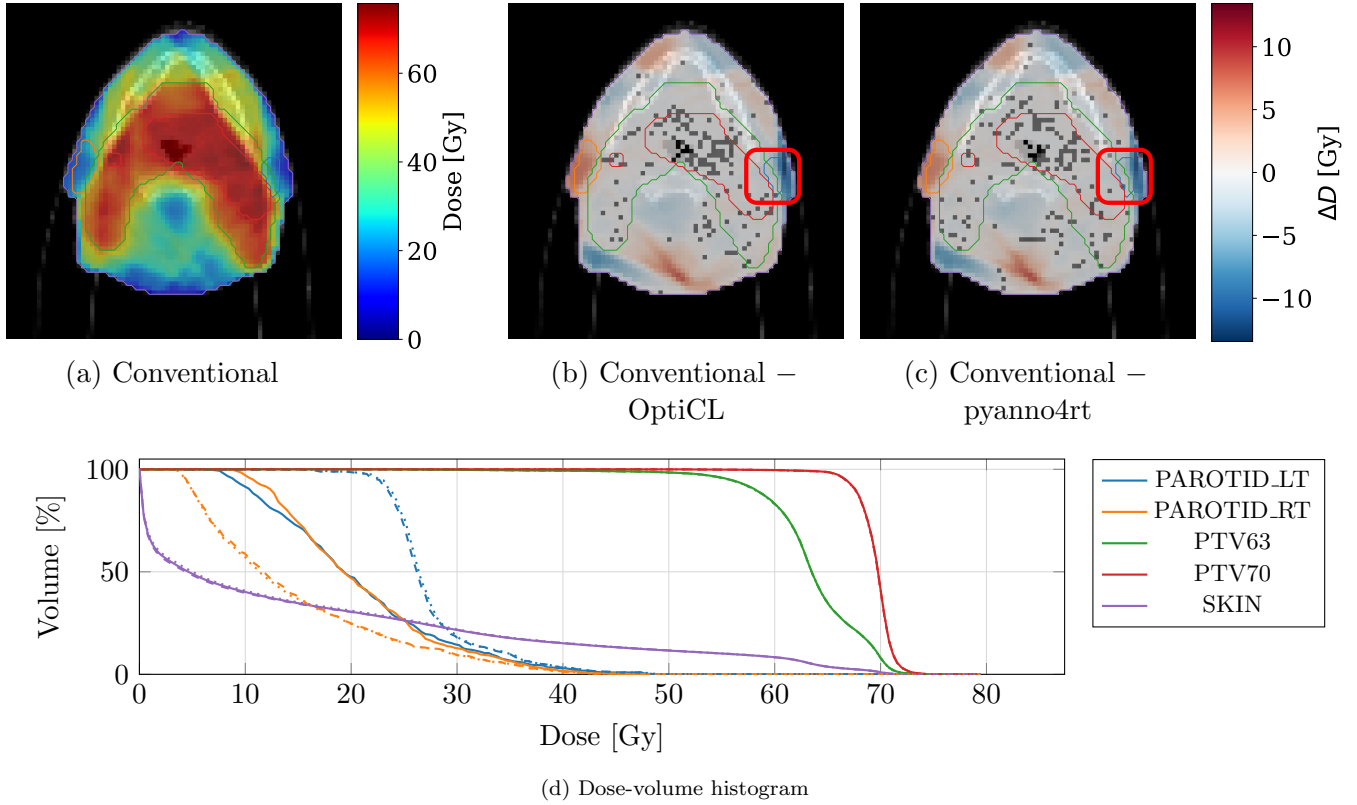}
    \caption{\textbf{Comparison of \enquote{bad} logistic regression-based example plans}. For HN, dose distributions from OptiCL (b) and pyanno4rt (c) are shown as dose differences relative to the conventional plan (a). DVHs (d) compare the conventional (\protect\solid), OptiCL (\protect\dashed), and pyanno4rt (\protect\dotted) plans. Both frameworks increase the mean dose to the left parotid gland beyond the recommended limit of \SI{25}{\gray} (red bounding box). This is driven by a logistic regression model coefficient of -0.04, which implies a negative association between mean dose and predicted risk of grade 2+ long-term xerostomia. The predicted NTCP decreases from \SI{48.2}{\percent} for the conventional plan to \SI{35.5}{\percent} for both model-based plans. In both cases, the dose difference $\Delta{D}$ is normalized to the prescribed dose $D_{\text{prescr}}$.}
    \label{fig:lr-bad-comparison}
\end{figure}

This vulnerability is demonstrated by the example in \Cref{fig:lr-bad-comparison}, where the underlying logistic regression model captured a negative correlation between the mean dose to the left parotid gland and the predicted risk of grade 2+ long-term xerostomia, reflected by a negative model coefficient of -0.04. Consequently, both MIP and NLP reward local dose escalation beyond the recommended limit of \SI{25}{\gray}. 

\subsection{Reproducibility} \label{subsec:disc-reproducibility}
An important contribution of this work is that the entire comparison was conducted using open-source software. OptiCL and pyanno4rt provide transparent implementations of their respective optimization paradigms, while matRad serves as a well-established and reproducible basis for dose calculation and plan generation \cite{OptiCL2021,pyanno4rt2026,Wieser2017}. This transparency is particularly relevant in ML-based treatment planning, where methodological claims are often difficult to verify without access to the underlying optimization pipeline. By utilizing open tools, this work facilitates independent reproduction, critical inspection, and further extension by the research community \cite{Kalendralis2021}.

Beyond transparency, the reusable interface developed to translate pyanno4rt models into OptiCL-compatible formulations provides a robust foundation for future benchmarking. This could support more systematic comparisons across diverse optimization models and TCP/NTCP architectures, potentially advancing the field toward shared benchmark problems for outcome-driven inverse planning. Standardized benchmarks, though currently lacking, would be highly valuable for disentangling advancements in predictive modeling from innovations in optimization methodology.

\subsection{Limitations}
\label{subsec:disc-limitations}
Several limitations of this study warrant consideration. 

First, the embedded ML models were trained on relatively small, NTCP-only datasets and did not consistently achieve high predictive performance. Accordingly, our findings should be interpreted primarily as a methodological comparison of optimization paradigms rather than as a clinical validation of the decision-support models. Moreover, tumor control was represented exclusively through dosimetric target objectives, without incorporating a TCP model. Consequently, the trade-off between predicted tumor control and predicted toxicity, which is central to outcome-based planning in its broader sense, was not investigated. Extending the proposed frameworks to incorporate TCP models and to jointly optimize TCP and NTCP represents a natural next step. This extension is technically feasible within the existing frameworks, provided that a suitable TCP or other clinically relevant outcome model (e.g., predicting progression-free survival) is available. 

Second, the evaluation was restricted to specific types of ML-based NTCP models, omitting, e.g., more intricate neural network architectures, nonlinear SVMs, ensemble models, and Bayesian networks; furthermore, the cohort was limited to eight HN and seven lung patients, which is sufficient to identify consistent trends but not to fully capture the broad anatomical variability encountered in clinical routine.

Third, the current planning framework does not yet include robust optimization or uncertainty-aware outcome modeling, both of which are essential for the safe practical implementation of ML-integrated treatment planning workflows. Other systematic errors are inherent to the pencil-beam dose calculation algorithms used. While these do not affect the comparison between the two frameworks, systematic dose errors and changes in dose calculation algorithms complicate consistent NTCP modeling with historical cohorts. In this case, the application of pencil-beam algorithms is deemed appropriate based on the provenance of the training data.

Fourth, the NTCP models used for plan evaluation are the same models embedded in the optimization process. Consequently, the reported NTCP differences reflect how effectively each framework minimizes the respective model predictions, rather than providing an independent assessment of plan quality. 

Finally, the hybridization analysis was intentionally limited to a single decision tree-based HN example and constitutes only a proof of concept. Combined with the observed scalability challenges, evidenced by OptiCL runs failing to terminate within the allotted optimization time, further underscore the necessity of the proposed hybrid strategies.

Overall, these findings support a nuanced perspective on ML-based inverse planning, suggesting that global constraint learning and nonlinear local optimization should be viewed as complementary rather than competing paradigms. NLP provides the computational efficiency and flexibility required for practical experimentation with complex differentiable outcome models, whereas MIP offers a rigorous path toward global optimality and explicit reasoning within discrete model structures. 

\section{Conclusion} \label{sec:conclusion}
In conclusion, this work addresses the current lack of standardized methodologies for integrating ML-based NTCP prediction models directly into treatment plan optimization. By evaluating the open-source frameworks OptiCL and pyanno4rt, we have demonstrated that both constraint-based and gradient-based strategies yield broadly consistent dose distributions and NTCP values across diverse clinical cases.

Our analysis highlights a fundamental trade-off: while OptiCL leverages linear and mixed-integer programming to achieve global optimality, pyanno4rt enables fast, gradient-based methods essential for efficient clinical iterations. Furthermore, our proposed hybridization strategies---exemplified by node pruning in decision trees---demonstrate that optimization model complexity can be scaled down without compromising dosimetric performance. These findings suggest that such hybrid approaches could bridge the gap between high-dimensional ML models and rigorous mathematical optimization, ultimately rendering MIP formulations feasible for large-scale clinical applications. This synergy paves the way for biologically individualized, ML-guided treatment planning in radiotherapy.

\section*{Conflict of Interest Statement}

The authors declare that the research was conducted in the absence of any commercial or financial relationships that could be construed as a potential conflict of interest.

\section*{ORCID iDs}
Tim Ortkamp \orcidlink{0009-0005-6603-3088},
Niklas Wahl \orcidlink{0000-0002-1451-223X}, Ali Ajdari \orcidlink{0000-0003-4672-4998}

\section*{Author Contributions}

\textbf{TO:} Conceptualization, Formal analysis, Methodology, Software, Writing – original draft, Writing – review \& editing; \textbf{FL:} Formal analysis, Software, Visualization, Writing – review \& editing; \textbf{ZL:} Resources, Writing - review \& editing; \textbf{NW:} Conceptualization, Methodology, Supervision, Writing – review \& editing; \textbf{AA:} Conceptualization, Methodology, Supervision, Writing – review \& editing \\

\noindent
All authors contributed to manuscript revision, read, and approved the submitted version.

\section*{Funding}
NW acknowledges funding by the Deutsche Forschungsgemeinschaft
(DFG, German Research Foundation) – Project No.~443188743.

\section*{Acknowledgments}
The present contribution is supported by the Helmholtz Association under the joint research school HIDSS4Health — Helmholtz Information and Data Science School for Health.

\section*{Data Availability Statement}

The datasets analyzed for this study are not publicly available due to institutional data protection policies. The data are hosted at the German Cancer Research Center (DKFZ) and Massachusetts General Hospital (MGH). Requests to access these datasets should be directed to Niklas Wahl, \href{mailto:n.wahl@dkfz-heidelberg.de}{n.wahl@dkfz-heidelberg.de}, and Ali Ajdari, \href{mailto:aajdari@mgh.harvard.edu}{aajdari@mgh.harvard.edu}.

\printbibliography

\end{document}